\documentstyle[aps]{revtex}

\input{psbox.tex}

\begin{document}

\title{Modeling premartensitic effects in Ni$_2$MnGa: A Mean Field and
Monte Carlo Simulation Study}

\author{Teresa Cast\'an and Eduard Vives}

\address{Departament d'Estructura i Constituents de la Mat\`{e}ria, Facultat 
de
F\'{\i}sica, \\Universitat de Barcelona, Diagonal 647, E-08028 Barcelona,
Catalonia}

\author{Per-Anker Lindg{\aa}rd}

\address{Condensed Matter Physics and Chemistry Department, Ris{\o} National 
Laboratory,
DK-4000 Roskilde, Denmark}

\maketitle

\begin{abstract}

The degenerate Blume-Emery-Griffiths model for martensitic transformations is
extended by including both structural and magnetic degrees of freedom in order
to elucidate premartensitic effects.  Special attention is paied to the effect
of the magnetoelastic coupling in Ni$_2$MnGa.  The microscopic model is
constructed and justified based on the analysis of the experimentally observed
strain variables and precursor phenomena.  The description includes the (local)
tetragonal distortion, the amplitude of the plane-modulating strain, and the
magnetization.  The model is solved by means of mean-field theory and Monte
Carlo simulations.  This last technique reveals the crucial importance of
fluctuations in pretransitional effects.  The results show that a variety of
premartensitic effects may appear due to the magnetoelastic coupling.  In the
mean-field formulation this coupling is quadratic in both the modulation
amplitude and the magnetization.  For large values of the magnetoelastic
coupling parameter we find a premartensitic first-order transition line ending
in a critical point.  This critical point is responsible for the existence of
large premartensitic fluctuations which manifest as broad peaks in the specific
heat, not always associated with a true phase transition.  The main conclusion
is that premartensitic effects result from the interplay between the softness of
the anomalous phonon driving the modulation and the magnetoelastic coupling.
In particular, the premartensitic transition occurs when such coupling is strong enough to
freeze the involved mode phonon.  The implication of the results in relation to
the available experimental data is discussed.

\end{abstract}


\newpage

\section{Introduction}

Many metals and alloys undergo a so called martensitic transition (MT) from an
open cubic phase at high temperatures to a more closed packed phase at lower
temperatures \cite{Nishiyama}.  It is a displacive, diffusionless, first-order
phase transition, accompanied by incomplete softening of certain transverse
phonon modes.  For Zr, which belongs to an ideally simple class of martensitic
materials, the pure group IV-metals \cite{Stassis,Petry}, it was demonstrated
that the first-order character can be understood as an effect of a coupling
between two simultaneous strains:  an internal two-plane shuffle strain and a
uniform strain \cite{Lindgard86}.

A rich variety of precursor phenomena \cite{Krumhansl90} have been observed in
(weakly) first-order MT.  Some of them, as the intermediate tweed structures
\cite{Shapiro86,Oshima88}, are not common to all materials but others,
intimately related to the transition mechanism, are present in almost all bcc
systems studied so far.  The most significant is the anomalously low
$TA_{2}[110]$ phonon branch ($[110]$ propagation, $[1\bar{1}0]$ polarization]),
accompanied by a low value of the elastic constant
$C^{\prime}=(C_{11}-C_{12})/2$.  Moreover, both the phonon branch and the
corresponding elastic stiffness soften with temperature.  Recently, a lot of
interest has been focused on the intermetallic Ni-Mn-Ga alloy close to the
stoichiometric composition Ni$_{2}$MnGa.  It is the only known ferromagnetic fcc
Heusler alloy undergoing a MT on cooling.  Besides its theoretical interest, it
may be of technological importance too since it opens the possibility of
controlling its shape memory properties (intimately related to the MT) by
applying an external magnetic field \cite{Vasilev99}.  It has a transition from
a bcc (neglecting the atomic order) to a low temperature tetragonal phase bct,
which is modulated by a five-plane shuffle strain \cite{Martynov92,Fritsch94}.
Particularly intriguing is that for nearly stoichiometric composition the full
MT is preceded by an intermediate phase in which apparently only the shuffle
strain is activated, but not the tetragonal strain
\cite{Planes97,Manosa97,Zheludev95,Zheludev96}.  This intermediate phase
consists in a micromodulated domain structure, without resulting macroscopic
tetragonal deformation so that the cubic symmetry is preserved\cite{Zheludev96}.
This is accompanied by a significant, although not complete, softening of the
$TA_2$ phonon branch at a wave vector $\xi_0=0.33$.  Only at lower temperatures,
at the martensitic transition point, the homogeneous tetragonal strain is
activated (and the modulation changes slightly).  This particular behavior
observed in Ni$_2$MnGa seems to be related to the influence of magnetism.  The
point is whether or not this intermediate modulated structure is a genuine
new phase.  Different behaviors have been reported in the literature.  For some
samples \cite{Planes97,Zheludev95,Zheludev96} there exists evidence for a true
phase transition of very weak first-order which is driven by a magnetoelastic
coupling.  The main proof of that is the fact that the intermediate transition
(IT) shifts with the external applied field \cite{Planes97} while no
(significant) magnetic field dependence has been found for the MT temperature
\cite{Obrado98,Zuo98}.  In other studies \cite{Stuhr97}, the authors could not
find any indication for an IT although precursors, clearly related to the
magnetization of the sample, have been observed.  Apparently, the only relevant
difference in the samples used by Planes {\it et al} \cite{Planes97}, Zheludev
{\it et al} \cite{Zheludev95,Zheludev96} and Stuhr {\it et al} \cite{Stuhr97} is
the content in Mn.  Very recently, it has been suggested \cite{Vasilev99} that
the tetragonal phase can be suppressed by increasing the concentration of Ni at
expenses of the content of Mn.  Moreover, the magnetoelastic effects have been
confirmed from the experimentally observed dependence of the elastic constants
on an external magnetic field \cite{Alfons99}.  In spite of the experimental
evidence for the magnetoelastic coupling in Ni$_2$MnGa, its microscopic origin
has not been established yet.

In the present paper we have theoretically investigated the nature of the bcc
to bct transition and constructed a model in order to solve the puzzling
behavior and to elucidate the role of the magnetic coupling, which
counter-intuitively seems to couple the ferromagnetic order stronger to the
modulating strain ($\eta$) than to the uniform tetragonal strain ($\epsilon$).
We shall demonstrate that the situation can be described by the degenerate
Blume-Emery-Griffiths model (DBEG) \cite{Vives96} extended to include coupling
to magnetic degrees of freedom, and with a interpretation of the variables,
appropriate to the present case.

The plan of the paper is the following:  First, in section II, we provide an
analysis of the experimental facts and a theoretical explanation of the observed
phenomena, which is used to justify a microscopic model presented in section III.  The
model is studied using first mean-field theory (section IV) and next Monte Carlo
simulation (section V).  The discussion from the comparative study is presented
in section VI where we provide also our conclusions taking into account the available
experimental data.

\section{Experimental facts and theoretical interpretation}

\subsection{Experimental Facts}

The structural properties of Ni$_2$MnGa have been investigated in a series of
papers 
\cite{Martynov92,Fritsch94,Planes97,Manosa97,Zheludev95,Zheludev96,Stuhr97,Koko96}.  
At 
high temperatures the alloy has the fcc ($L2_{1}$) Heusler structure
which, neglecting the atomic order, can be regarded as a bcc lattice.  It is
paramagnetic at high temperatures with the magnetic moments mainly on the Mn
sites \cite{Webster84}.  At temperatures below $T_m$, it orders
ferromagnetically with no particular easy direction of the moment.  At the
temperature $T_M$ ($<T_m$) there is a (first-order) structural phase transition
of the martensitic type to an {\it average} tetragonal structure, which
additionally is modulated by a transverse five-layer shuffling strain.  Prior to
this transition precursor structures of that phase as well as of the fcc having
a six-plane modulation have been observed in neutron scattering experiments
\cite{Zheludev96,Stuhr97}.  This may happen also as a transition (first-order)
at a temperature $T_I$ ($T_M<T_I<T_m$) giving rise to a genuine intermediate
phase \cite{PMT} without any macroscopic tetragonal deformation
\cite{Zheludev96}.

The above temperatures $T_m$, $T_M$ and $T_I$ are extremely sensitive to the
composition and atomic ordering of the sample.  Thus $T_m$ may vary between 360K
to 395K, whereas $T_M$ may vary from 175K to 450K.  In the sample studied by
Stuhr {\it et al.}  \cite{Stuhr97}, $T_m\approx 364$K and $T_M\approx 284$K with
precursor signs for $T>T_M$, but no intermediate phase was found.  In the sample
studied by Zheludev {\it et al.}  \cite{Zheludev95,Zheludev96} $T_m \approx
380$K and $T_M\approx 220$K and an {\it intermediate phase} is observed below
$T_I \approx 260$K.  Similarly, Planes {\it et al} \cite{Planes97} studied a
sample with $T_m\approx 381$K and $T_M \approx 175$K and found an intermediate
transition at $T_I \approx 230$K.  In spite of the large variation in the
temperatures, we assume the basic physics is the same, and hence we shall use
the information obtained by observations on different samples.  In particular,
we shall attempt to explain the precursor and the intermediate phase phenomena
by constructing an effective microscopic Hamiltonian, which allows analysis
beyond a mean-field or Landau expansion treatments
\cite{Vasilev99,Planes97,Lvov98}.

\subsection{The two-strain model}

Although Ni$_2$MnGa is a metallic alloy for which both the structure and the
magnetism is determined by the conduction electrons, it is instructive to
consider a model system for the mismatch between cubic crystals, similarly as it
is found in KCl grown on NaCl (001) \cite{Baker1,Baker2}.  In this case, the
very large ($\sim 17\%$) lattice constant mismatch makes the interface to buckle
simultaneously in the [110] and $[1\bar{1}0]$ directions.  This produces
superstructures consisting of seven NaCl and six KBr layers, or multipla
thereof, perpendicular to the interface, while preserving the square symmetry.
Both the NaCl and the KBr crystals are modulated since they have similar elastic
properties.  This gives rise to superstructure peaks in the X-ray and Helium
scattering spectra, precisely as those observed in the intermediate phase of
Ni$_2$MnGa, where one also would expect more higher order satellites in
off-symmetry directions to be found, but not so far looked for experimentally.
The buckling gives rise to a variation of the local [001] direction and a change
in lattice constant perpendicular to the interface (corresponding to a local
tetragonal distortion).  We stress that for the ionic crystals the forces by no
means triggers the local modulating strains, which are caused by the forced
contact between unequal crystals.  For Ni$_2$MnGa the situation is rather
different.  Here, a nesting feature of the Fermi surface causes a strong
electron-phonon coupling and an incipient soft phonon mode at ${\vec q} =
\langle \xi,\xi,0 \rangle$ positions in the fcc phase, where $\xi\sim
\frac{1}{3}\,$ \cite{Zheludev96,Stuhr97}.  This is presumably the driving
mechanism in Ni$_2$MnGa, and giving, as a consequence, the tetragonal
distortion.  As precursor phenomena, quasi elastic peaks are observed at
$\xi=\frac{1}{3}$ and $\xi=\frac{1}{6}$ \cite{Stuhr97} corresponding to a
similar six-plane modulation as discussed for the alkali salt interface.  To
match this to the fcc (001) lattice plane it is advantageous to make a lattice
mismatch and expand the lattice and to rumple the interface, or in other words
making the [001] direction fluctuate in epitaxial-like angles.  Stuhr {\it et
al} \cite{Stuhr97} have shown that the six-plane fcc modulation corresponds
precisely to a five-plane modulation of the tetragonal phase, and the latter is
found as a precursor phenomenon.  An 5:6 expansion would be very drastic.
However it suffices to create a superstructure of a common divisor, for example
a 30 or 60 plane repeat distance.  The latter would correspond to a lattice
mismatch of 1.67\%, which is very close to the mismatch observed between the fcc
and tetragonal phase of Ni$_2$MnGa:  $\sim1.6\%$ \cite{Webster84}.  Hence we
argue that the electronically driven six-plane modulation in turn also causes
the tetragonal distortion.  The theory for why it is advantageous for mismatched
crystals at epitaxial interfaces to develop mutual superstructure peaks was
discussed in more detail by Vives and Lindg\aa rd \cite{Vives91}.  The
modulation occurs simultaneously in the ${\vec q}=[\frac{1}{3},\frac{1}{3},0]$
and ${\vec q}=[\frac{1}{3},-\frac{1}{3},0]$ directions, thus preserving the
square symmetry of the (001) plane and yielding a modulated (001) {\it plane},
rather than a {\it direction}.  In the fcc structure there are three such
equivalent planes (100), (010) and (001).  In order to (essentially) preserve
the volume, $V$, a change of which is expensive for the electrons, the c-axis
[001] must shrink (from $T=4.2$ K to $295$ K it is indeed observed that
$V^\frac{1}{3}$ only increases by 0.5\% \cite{Webster84}).  Therefore a model
must include a coupling between the {\it plane} modulation $\eta$ and the
tetragonal strain $\epsilon$.  Then, the question is:  how can there be apparent
separate temperatures for the onset of ordering of the two kind of strains?
This is possible if in the intermediate phase the tetragonal strain is only
local, varying in direction and also in the allowed directions in the fcc
crystal.  Moreover, only in the martensitic phase microcrystals with a resulting
tetragonal strain should be formed.  The corresponding tetragonal structure is
observed as highly mosaic with a large variation of the [001] directions
\cite{Stuhr97}, in agreement with the above picture.

\subsection{The Landau models}

Very recently, Landau models for the MT in cubic ferromagnetic materials have
been proposed \cite{Vasilev99,Lvov98}.  In these models, the magnetoelastic
coupling between the uniform strain tensor and the (vector) magnetization is
fully considered.  Nevertheless, they seem more appropriated to the study of the
magnetic properties of the martensitic phase rather than to the analysis of the
IT itself.  Here, accordingly with the discussion given above, we shall adopt a
different strategy and study the structural transition in Ni$_2$MnGa in terms of
a Landau expansion of the only most relevant strain and magnetization variables,
including non-uniform strains or modulations.  Then, similarly to the case for
the bcc to hcp transition in Zr \cite{Lindgard86}, we include the following
terms, which are allowed by symmetry:
\begin{eqnarray}
{\cal F}(\eta,\epsilon)&=&\frac{1}{2}\,\mu\,       \omega_s^2\;
\eta^2+\frac{1}{4}\,B\;\eta^4  +     \frac{1}{6}\,  C\;   \eta^6\cdots
\nonumber \\ &+&  \frac{1}{2}\, C'\;\epsilon^2+\cdots \nonumber \\ &+&
\Delta\; \epsilon\; \eta^2,
\label{eq1}
\end{eqnarray} 
where $\eta$ is the discussed plane modulation strain and $\epsilon$ the {\it
local} uniform contraction perpendicular to that plane, but we do not consider
higher order uniform strain terms.  Here $\omega_s^2 = a(T-T_s)$ is the squared
frequency \cite{comment} for the incipient soft mode phonon with ${\vec q}
=\langle \frac{1}{3},\frac{1}{3},0\rangle$, other constants are positive, and
$C'=\frac{1}{2}(C_{\rm 11} - C_{\rm 12})$, which is small and temperature
dependent.  By eliminating the local tetragonal strain $\epsilon$ we can
\cite{Lindgard86}, write the free-energy along the optimum energy path involving
both $\epsilon=-\Delta^2/\,C' \; \eta^2$ and $\eta$ as
\begin{eqnarray}
{\cal F}(\tilde{\eta}) &=& \frac{1}{2}\,\mu  \,   \omega_s^2\;   \tilde{\eta}^2+
\frac{1}{4}\,\tilde{B}\;\tilde{\eta}^4       +    \frac{1}{6}\,    C\;
\tilde{\eta}^6\cdots,\nonumber \\ \tilde{B}&=&B-2\Delta^2/C'.
\label{eq2}
\end{eqnarray}
The coupling between the two strains therefore makes it possible for $\tilde{B}$
to become negative and hence to cause a first-order transition at $T_M$ before
the soft mode transition at $T=T_s$, even without the coupling with the
magnetism.

Next let us consider the influence of the magnetism.  A ferromagnetic moment
$\vec{m}$ can influence an itinerant magnet as $Ni_2 Mn Ga$ in two ways.
Firstly, giving rise to a splitting of the electronic energy bands, which is
proportional to the amplitude $|m|$, but which is not sensitive to the moment
direction.  This will cause a change in the soft phonon frequency below $T_m$
proportional to $|m|^2$:
\begin{eqnarray} 
\omega_s^2(m)= a(T-T_s) - u |m|^2 = a(T-T_s)+b(T-T_m),
\label{eq3}
\end{eqnarray}
where $a$, $b$ and $u$ are positive constants, and we have assumed a mean-field
behavior for $|m|^2$.  A measurement of the soft mode frequency-squared should
therefore exhibit a kink at the magnetic ordering temperature.  This is exactly
the behavior observed by Stuhr {\it et al} \cite{Stuhr97}, who found $a=0.018$
meV$^2$/K and $b=0.020$ meV$^2$/K.  Unfortunately for the samples showing the IT
\cite{Planes97,Zheludev95,Zheludev96} measurements in the paramagnetic phase are
not available.  Nevertheless, in what follows, we shall assume the behavior
discussed above for all the samples.  Hence the magnetic free-energy part can be
written as:
\begin{eqnarray}
{\cal F}_{\rm mag}(m,\tilde{\eta})=  \frac{1}{2} \,  \alpha\; |m|^2 +  
\frac{1}{4}\,
\beta\;  |m|^4     +  \cdots  +\frac{1}{2}\,\gamma\;    |m|^2  \tilde{\eta}^2,
\label{eq4}
\end{eqnarray}
where $\alpha=A(T_c-T)$ and $T_c$ is the magnetic transition in absence of
magnetoelastic coupling.  $A$ and $\beta$ are positive parameters and $\gamma=-
\mu u$ is yielding the coupling between the amplitude-squared of the
magnetization and the effective modulating strain.  By eliminating $|m|^2$ one
can write an effective free-energy in the form of the eq.(\ref{eq2}).  This
yields a further temperature dependence of $\omega_s^2$ and $\tilde{B}$, as
discussed by Planes {\it et al} \cite{Planes97}.

The other possible coupling between the magnetization and the structure is by
magnetostriction, which deforms the crystal in the {\it direction} of the
magnetization \cite{Vasilev99}.  Experimentally, the easy direction of
magnetization is not known with certainty, even in the tetragonal phase.
Webster {\it et al} \cite{Webster84} propose it might be in the $\langle 111
\rangle$ directions of the $L2_1$ phase \cite{mdirec}, but that other directions
are almost as likely.  Analysing their data perhaps allows the conclusion that
it is at least {\it not} in the tetragonal [001] direction, which would normally
have been the obvious choice.  If so, it is confined to the (001) plane, say,
which has four-fold symmetry and therefore not yielding a strong easy axis.  The
lowest order coupling would then be of the form $ +\vec{m}^2 \epsilon^2$.  An
interesting possibility is if the easy direction is along the [100] direction,
because this may belong to two different modulation planes, (001) and (010),
therefore yielding a minimum magnetostriction because it cannot distinguish
between a [001] and a [010] tetragonal strain.  The effect is further reduced
because of the equivalence between moments in the [100] and [010] directions in
the modulated plane (001).  Inclusion of the coupling can be done in the
eq.({\ref{eq4}), with no qualitative change (except for an induced dependence on
the magnetization of $C'$ \cite{Planes97}), and hence we shall neglect it in the
following.  If the coupling is sufficiently strong, it would no doubt prevent
the existence of an intermediate phase in Ni$_2$MnGa.

The conclusion of the above analysis is that the phases may be characterized by
the minimum path modulation strain $\tilde{\eta}$ in eq.(\ref{eq2}) which
includes a finite, but local tetragonal strain which is on average zero because
of fluctuating directions.  This is consistent with the observation of a
significant broadening of the fcc (002) peak in the neutron scattering data
\cite{Stuhr97}.  This strain is coupled to the magnetization such that a
ferromagnetic moment on neighboring sites will favor a modulation also at these
sites, due to a change in the local band structure, but without differentiating
between the three possible planar modulations and accompanying tetragonal local
strains.  In order to give meaning to the notion of a local band structure, we
consider sites not as the Mn atom positions, but at a more coarse grained level,
{\it e.g.  } at the unit cell level, say.  At the MT a resulting tetragonal
strain, $\epsilon$, may arise as the interfaces between the modulated fcc and
the modulated and locally tetragonal crystallites grow larger and thicker.  This
strain is on the average in the $\langle 001 \rangle$ directions but with
considerable variations in the epitaxial angles relative to those.  To construct
a simplified model, we shall consider one with only two variables such that each
has only two degrees of freedom.  One variable with only two values is
representing the plane modulations $\tilde{\eta}$, and the other also with only
two values representing the resulting tetragonal strain $\epsilon$.  This does
not correspond to considering the behavior of a subspace of variants, but
constitutes a simplified statistical analogue to the physics of Ni$_2$MnGa.

\section{The Model}

The desired degrees of freedom can be represented by the $p$-degenerate
Blume-Emery-Griffiths Hamiltonian (DBEG)\cite{Vives96}.  It was first introduced
with the aim to account for the entropy stabilization of the high temperature
phase in martensitic transitions.  Recently, it has been shown \cite{Burkhardt}
that it is equivalent, with respect to universal properties, to that of the
ordinary BEG model with the crystal field shifted by a term $k_B T \ln p$.

Although the DBEG model is very simple it includes most of the relevant physical
ingredients to understand MT, namely a multivariant low temperature deformed
phase and a high temperature {\it average} cubic phase with enhanced entropy.
The model is an extension of the ordinary three-state Blume-Emery-Griffiths
Hamiltonian defined on a lattice, which we shall take as simple cubic (or
square), as motivated above.  On each lattice site $i=1,..,N$ a variable
$\sigma_i = 1,0,-1$ represents the deformation state near each site on the
lattice.  The state $\sigma_i=0$ represents the undistorted phase, and it is
chosen to be $p$-fold degenerate ($p\ge 1$), in order to approximately account
for the high entropy of vibration of the cubic phase.  The states $\sigma_i=\pm
1$ represent the distorted phase.  The Hamiltonian accounting for the energy
gain in having the same structure on neighboring sites was written as
\cite{Vives96}
\begin{equation} 
{\cal H}_M =  -  J \sum_{\langle  i,j \rangle}^{n.n.}   \sigma_i \sigma_j - K
\sum_{\langle i,j \rangle}^{n.n.}  (1-\sigma_i^2)(1-\sigma_j^2).
\label{DBEG} 
\end{equation}
where the sums are performed over all nearest-neighbor pairs.  In what follows
we will take $J>0$ as the unit of energy and work in terms of reduced magnitudes
defined as ${\cal H}^* = {\cal H}/J$, $K^* = K/J \dots$.

Two order parameters can be defined, $\phi_1= \sum \sigma_i /N $ and
$\phi_2=\sum \sigma_i^2/N$.  The model in (\ref{DBEG}) was solved, for $K^* \geq
0$, by mean-field and Monte Carlo simulation techniques \cite{Vives96}.  We
found a phase transition from a cubic (disordered) phase with $\phi_1=0$ to a
tetragonal (ordered) phase with $\phi_1 \neq 0$.  The behavior of the secondary
order parameter $\phi_2$ in the disordered phase depends on $K^*$ and $T^*$, but
exhibits non-analytical behavior at the same temperature as $\phi_1$.  Large
values of $K^*$ stabilize the cubic phase.  Moreover, the $K^*$ parameter and
the degeneration $p$ of the $0$-state control the order of the transition, which
changes from being of second order (for low values of $K^*$) to first-order.

The DBEG model \cite{Vives96} was introduced as a simplification and a further
generalization of the Lindg{\aa}rd-Mouritsen model \cite{Lindgard86}, initially
designed to mimic the bcc to hcp transition in Zr.  Consequently, it naturally
inherits the identification of the two order parameters suitable for Zr:
$\phi_1$ is the two-plane shuffle strain and $\phi_2$ the homogeneous strain.
In the previous work \cite{Vives96} this was not emphasized since both $\phi_1$
and $\phi_2$ exhibit a phase transition at the same point.

For Ni$_2$MnGa it is more natural to identify the order parameters in the
opposite way.  Therefore, in the present work, we define:
\begin{equation}
\epsilon=\frac{\sum \sigma_i}{N},
\end{equation}
\begin{equation}
q=\frac{\sum \sigma_i^2}{N}.
\end{equation}
The breaking symmetry order parameter $\epsilon$ corresponds to the tetragonal
distorsion in the sense that if $\epsilon=0$ (equal population of $\sigma_i=+1$
and $-1$) it corresponds to having all the variants equally populated, and hence
an {\it average} cubic phase.  For $q$ the relation is more involved because of
the complicated physics of the modulating strain $\tilde{\eta}$.  In the high
temperature cubic phase, when the $\sigma_i$ variables distribute at random $q=
q_0 = 2 / (p+2)$.  Let us assign the difference $q_0 - q$ with the amplitude
$\tilde{\eta}$ of the plane modulating strain without distinguishing between the
three possible modulating planes.

We now include the magnetic degrees of freedom by means of spin variables
$S_i=\pm 1$ (defined on the lattice site $i=1,..,N$) having a ferromagnetic
Ising interaction.  Thus, the purely magnetic contribution is:
\begin{equation} 
{\cal H}_m^*= -J^*_m \sum_{\langle i,j \rangle}\; S_i\, S_j,
\label{Hmag} 
\end{equation}
where $J^*_m >0$.

The total Hamiltonian should further include a coupling term between the
structural and magnetic variables.  We have argued in section II that the
magnetic influence of electronic properties gives rise to a coupling between the
magnetic moment and the plane modulation.  On a microscopic level let us assume
that the presence of a moment of neighboring sites gives rise to a modulation on
neighboring sites as motivated previously.  To describe this, let us consider
the following symmetry allowed \cite{magnetostriction} interaction
contributions:
\begin{eqnarray} 
{\cal H}_{\rm   int}^* = &-& U^*_{11} \sum_{\langle i,j \rangle}\; S_i\,S_j\; 
\sigma_i^2\,\sigma_j^2 \nonumber \\ &-& U^*_{00} \sum_{\langle i,j \rangle}\;  
S_i\,S_j\;
(1-\sigma_i^2)(1-\sigma_j^2) \nonumber \\ &-& U^*_{10}  \sum_{\langle  i,j   
\rangle}\;  S_i\,S_j\;
\,[\sigma_i^2(1 - \sigma_j^2)+ \sigma_j^2 (1 - \sigma_i^2)],
\label{eq7}
\end{eqnarray}
that may be rewritten as:
\begin{eqnarray}
{\cal H}_{\rm int}^* = &-& (U^*_{11} +U^*_{00} - 2 U^*_{10}) \sum_{\langle i,j 
\rangle}\; S_i\,S_j\; \left (\frac{1}{2} - \sigma_i^2 \right ) \left (\frac{1}{2} - 
\sigma_j^2 \right ) 
\nonumber \\ 
&+& \frac{1}{2}(U^*_{11} - U^*_{00}) \sum_{\langle i,j \rangle}\; S_i\,S_j\;   
\left [ \left (\frac{1}{2}-\sigma_i^2 \right ) + 
\left (\frac{1}{2}-\sigma_j^2 \right ) \right ] \nonumber \\ 
&-&  \frac{1}{4} (U^*_{11}+ U^*_{00} +2 U^*_{10}) \sum_{\langle i,j \rangle}\; 
S_i\,S_j. 
\label{Hint1}
\end{eqnarray}
In the pure ferromagnetic phase ($S_i=1$) this Hamiltonian can be viewed as an
Ising Model for the variables $1/2-\sigma_i^2$.  Thus a phase transition exists
for $(U^*_{11}+U^*_{00}-2 U^*_{10})>0$ and $U^*_{11}$ = $U^*_{00}$.  For
simplicity, in what follows we shall take $U^*_{11}$ = $U^*_{00}=0$ and denote
$U^*= U^*_{10}<0$.  Then, the coupling Hamiltonian becomes:
\begin{eqnarray} 
{\cal H}_{\rm int}^* = 2 U^* \sum_{\langle i,j \rangle}\;   S_i\,S_j\;
\left (\frac{1}{2} - \sigma_j^2 \right )   \left (\frac{1}{2} - \sigma_i^2 \right ) - 
\frac{1}{2} U^*
\sum_{\langle i,j \rangle}\; S_i\,S_j.
\label{Hint2}
\end{eqnarray}
As we shall see in the next section, in the mean-field approximation, the first
term becomes of the form of the coupling term in eq.({\ref{eq4}) whereas the
last term gives a simple modification of the purely magnetic interaction
$J^{*}_m$ defined in eq.(\ref{Hmag}).  Furthermore, (\ref{Hint2}) shows that it
may be particularly convenient to choose $p=2$, which gives $q_0=1/2$.

The total Hamiltonian model for Ni$_2$MnGa can then be written as
\begin{equation} 
{\cal H}^* = {\cal H}_M^* + {\cal H}_m^* + {\cal H}_{\rm int}^*,
\label{model} 
\end{equation}
with ${\cal H}^*_M$, ${\cal H}^*_m$ and ${\cal H}^*_{int}$ respectively given by
expressions (\ref{DBEG}), (\ref{Hmag}) and (\ref{Hint2}).  We shall demonstrate
that it is possible to split up the structural transition into one determined by
the order parameter $q$, which we will associate with the IT and another one,
determined by $\epsilon$, to be associated with the tetragonal deformation
occurring at the MT.

\section{Mean-Field Treatment}

In this section we solve the presented model (\ref{model}) by using standard
mean-field techniques.  The state of order of the system depends on the
occupation numbers $N_{\sigma}^{S}$.  This stands for the number of points in
the structural $\sigma={-1,+1,0}$ and magnetic $S={+,-}$ state.  There are six
different occupation numbers which should fulfil the following normalization
condition:
\begin{equation}
N_1^+ + N_1^- + N_0^+ + N_0^- + N_{-1}^+ + N_{-1}^- = N,
\end{equation}
where  $N$ is the total  number of points in  the lattice.  We define the
following order parameters:
\begin{eqnarray} 
N \epsilon &=& \sum_i \sigma_i =  (N_{1}^{+} + N_{1}^{-}) -(N_{-1}^{+}
+ N_{-1}^{-})  \\ N q   &=& \sum_i {\sigma_i}^2   =  N - (N_{0}^{+}  +
N_{0}^{-})\\ N m &=& \sum_i S_i = N_{1}^{+} + N_{-1}^{+} + N_{0}^{+} -
N_{1}^{-} -  N_{-1}^{-}   -  N_{0}^{-}\\ N  m_0    &=& \sum_i   (1   -
{\sigma_i}^2)S_i = N_{0}^{+}  -N_{0}^{-}\\  N m_1 &=&  \sum_i \sigma_i
S_i = (N_{1}^{+} - N_{1}^{-}) - (N_{-1}^{+} - N_{-1}^{-}).
\label{defordpar}
\end{eqnarray}
The corresponding entropy can be written as:
\begin{equation}
 S_{MF}/           k_B              =         \ln        \left       (
\frac{N!}{{N_{1}}^{+}!{N_{1}}^{-}!{N_{-1}}^{+}!
{N_{-1}}^{-}!{N_{0}}^{+}!  {N_{0}}^{-}!}  p^{(N_0^+ + N_0^-)} \right )
\end{equation}
where  $p  \geq 1$ is  the  degeneracy factor of   the $0$-state.  The
mean-field expression of the free-energy per particle is:
\begin{eqnarray}
{\cal F}_{MF}^* &=&  =  - \left  (  \epsilon^2 + K^*
(1-q)^2 +  J^*_m m^2 +2 U^*  m_0 (m - m_0) \right  )  \nonumber \\ &+&
\frac{T^*}{2}   \left    [   (q+\epsilon+m+m_1-m_0)     \ln   \left  (
\frac{q+\epsilon+m+m_1-m_0}{4}  \right )  \right  .   \nonumber \\ &+&
(q-\epsilon+m-m_1-m_0)    \ln   \left ( \frac{q-\epsilon+m-m_1-m_0}{4}
\right  )    \nonumber  \\  &+& (q+\epsilon-m-m_1+m_0)  \ln    \left (
\frac{q+\epsilon-m-m_1+m_0}{4} \right    )  \nonumber       \\     &+&
(q-\epsilon-m+m_1+m_0)  \ln  \left  (   \frac{q-\epsilon-m+m_1+m_0}{4}
\right )  \nonumber \\ &+& 2 (1-q+m_0)  \ln  \left ( \frac{1-q+m_0}{4}
\right  ) \nonumber \\ &+&  2 (1-q-m_0)  \ln \left ( \frac{1-q-m_0}{4}
\right ) \nonumber \\ &-& \left .  4 (1-q) \ln \left (\frac{p}{2}\right)
\right ],
\end{eqnarray}
where $T^* = k_B T/zJ$, and $z$ is the coordination number of a given site.
Notice that $m_1$ appears only in the entropic contribution to the free-energy.
Standard minimization with respect $m_1$ renders the following relationship
$m_1= \frac{\epsilon(m-m_0)}{q}$, which has to be fulfilled at all temperatures.
Then, after substitution in eq.(20) we obtain the following expression
for the free-energy as function of $\epsilon,q,m,m_0$:
\begin{eqnarray}
{\cal F}_{MF}^* &=& -(\epsilon^2 +K^* (1-q)^2 + J^*_m m^2 + 2 U^* m_0 (m
- m_0))    \nonumber   \\ &+&  T^*    \left [  (q+\epsilon)  \ln \left
(\frac{q+\epsilon}{2}     \right    )    +   (q-\epsilon)   \ln  \left
(\frac{q-\epsilon}{2} \right ) \right  .   \nonumber \\ &+&  (q+m-m_0)
\ln  \left   (\frac{q+m-m_0}{2} \right   )     + (q-m+m_0) \ln   \left
(\frac{q-m+m_0}{2} \right )  \nonumber   \\ &+& (1-q+m_0) \ln    \left
(\frac{1-q+m_0}{4} \right ) +  (1-q-m_0) \ln \left  (\frac{1-q-m_0}{4}
\right ) \nonumber \\ &-& \left .  2 (1-q) \ln \left( \frac{p}{2} \right)
- 2 q \ln q \right ].
\end{eqnarray}
Further  minimization with respect   the other four  order parameters
yields to the next set of coupled equations:
\begin{equation} 
\epsilon = \frac{T^*}{2} \ln \frac{q+\epsilon}{q-\epsilon},
\label{epsilon}
\end{equation}
\begin{equation} 
-\left[K^*(1-q) + T^* \ln(\frac{p}{2}) \right] = \frac{T^*}{2} \ln  \left
(\frac{(q+\epsilon)(q-\epsilon)(q+m-m_0)(q-m+m_0)}       {q^2
(1-q+m_0)(1-q-m_0)}  \right ),
\label{q}
\end{equation}
\begin{equation}
J^*_m m + U^* m_0 = \frac{T^*}{2} \ln \frac{q+m-m_0}{q-m+m_0},
 \label{m}
\end{equation}
\begin{equation}
U^*    (m-2  m_0) =   \frac{T^*}{2}  \ln  \frac {(q-m+m_0) (1-q+m_0)}
{(q+m-m_0)(1-q-m_0)}.
\label{m0}
\end{equation}
Their solution gives the temperature dependence of the order parameters.
Between all possible solutions only the absolute minima correspond to
thermodynamic equilibrium.  This requires the analysis of the second derivatives
of the free-energy.

The space of the model parameters of interest here is limited by the conditions
$J^*_m>0$, $U^*<0$ and $J^*_m+U^*>0$.  Then, for appropriated values of $K^*$,
there exists three phase transitions at the temperatures $T^*_M<T^*_I<T^*_m$
associated with $\epsilon$, $q$ and $m$ respectively.  In what follows we shall
fix the value of the magnetic interaction to $J^*_m=4.0$, thus determining the
distance between $T^*_M$ and $T^*_m$, and use different values of the coupling
parameter $U^*$ ($0<U^*<-J^*_m=-4$).  Then, the values of $K^*$ for which
the IT exists are determined by eq.  (\ref{q}).  Indeed, by setting
$\epsilon=0$, it follows that for $K^*(T^*)=-2T^* \ln (p/2)$ the order parameter
$q$ has a continuous phase transition.  The results will be presented for two
values of the degeneracy factor, $p=2$ and $p=4$.  In Table I we give the
identification of the different phases in relation to the problem of interest
here and their corresponding abbreviated notation.  Figures \ref{FIG1} and
\ref{FIG2} show the temperature behavior of the order parameters for a given
value of $U^*=-3.50$ along the path determined by the condition $K^*(T^*)=-2T^*
\ln (p/2)$ (coexistence line).  In particular, for $p=2$ one has $K^*(T^*)=0$.
For the sake of clarity, $m_0(T^*)$ is not shown.  At high temperature, a
magnetic transition appears at $T^*_m= J^*_m + U^*/2$ from a paramagnetic cubic
phase (PC) ($\epsilon=0,q=1/2,m=m_0=0$) to a ferromagnetic {\it pure} cubic
phase (FC) ($\epsilon=0,q=1/2,m=2m_0 \neq 0$).  From equation (\ref{q}) it is
easy to see that, for $q=1/2$ ($T^*< T^*_I$), $m$ and $m_0$ are not independent
but $m=2m_0$.  At lower temperatures, the order parameter $q$ separates into two
branches $q^+ (q>1/2)$ and $q^- (q<1/2)$.  This occurs at the critical point
$T^*_{Ic}$ and separates two different ferromagnetic phases both with
$\epsilon=0$ but with different values of $q$:  a {\it pure} cubic phase
($\epsilon=0,q=1/2,m=2m_0 \neq 0$) and an {\it average} cubic phase
($\epsilon=0,q \neq 1/2,m \neq m_0 \neq 0$).  The two branches of $q$ are
identified with the tetragonal-like modulation ($q^+$) and the fcc-like
modulation ($q^-$) in the sense that when all are equally populated the cubic
symmetry is preserved.  Finally, at a temperature $T^*_M$, a martensitic-like
transition to a ferromagnetic martensitic (tetragonal) phase (FMT) ($\epsilon
\neq 0,q=q^+,m \neq 0,m_0 \neq 0$) occurs.

In figures \ref{FIG3} and \ref{FIG4} we show the phase diagram as function of
$K^*$.  The intermediate transition (IT) is of first-order and it is represented
by the dashed line ($T^*_I$) separating, below the critical point (black
diamond), the regions with $q^+$ and $q^-$.  For $p=2$ (Fig.  \ref{FIG3}), this
boundary is a vertical straight line and therefore cannot be crossed by sweeping
$T^*$ at constant $K^*$.  We advance here that this is a mean-field artifact.
The exact treatment by Monte Carlo simulation will show that this transition
line is always bent, even for $p=2$.  The mean-field solution renders a real
first-order IT only for $p>2$, as it can be seen in Fig.  \ref{FIG4} for $p=4$.
As an example, in figure \ref{FIG5} we show the temperature behavior of the
order parameters for $p=4$ and three values of $K^*$ around the critical value
$K^*_c$; $K^*<K^*_c$ (a), $K^*=K^*_c$ (b) and $K^*>K^*_c$ (c).  The insets shows
an enlarged view of the magnetization behavior around $T^*_I$.

In figure \ref{FIG6} we show the location of the critical point ($T^*_{Ic}$) for
$p=2$ ($K^*_c=0$) with respect to the magnetic ($T^*_m$) and the martensitic
($T^*_M$) transitions as function of the coupling parameter $U^*$.  One observes
that, as the strength of the coupling decreases, the critical point approaches
the martensitic line in such a way that the IT disappears well before $U^*$
becomes zero.  This is consistent with some experimental observations
\cite{Alfons99} indicating that strong coupling is required for the IT to
appear.  The same behavior is found for $p=4$, as shown in Fig.  \ref{FIG7}. We note 
that, in this case, the critical points correspond to different values of
$K^*$ ($K^*_c=-2T^* \ln \frac {p}{2}$).  Figure \ref{FIG8} shows the location of
the critical points (thick line) and the region of first-order IT in the
$U^*$-$K^*$ plane.  For very weak coupling, the IT does not exists.  For $p=4$, as the
strength of the magnetoelastic coupling increases, the IT exists for a larger
interval of values of $K^*$ above $K^*_c$.

The magnetic transition is of second order for all values of model parameters
studied.  The MT is found to be first-order only for low values of the coupling
strength for which $\epsilon$ and $q$ order simultaneously.  Notice that this
does not contradict the results obtained previously for the DBEG with $U^*=0$
and $K^* \geq 0$.  Actually, for $p=4$, it was found that the MT transition is
discontinuous only for $K^*>0$ \cite{Vives96}.  Thus, the mean-field solution of
the present model does not reproduce the evident first-order character of the MT
in Ni$_2$MnGa.  At this point, this could be attributed to the insufficient
accuracy of the mean-field treatment but, in the next section, we will see that
Monte Carlo studies render the same feature.  Thus, it is more
plausible that it is due to the incompleteness of the model, addressed
preferently to the study of the IT and related effects.  Actually, our model
description is done in terms of a single modulation strain whereas it is known
\cite{Martynov95} that the MT and the IT phases have different modulations.

In conclusion, the main effect of the magnetoelastic coupling parameter $U^*$ is
to generate a critical point ($T^*_{Ic}$, $K^*_c=-2T^* \ln \frac {p}{2}$),
between the magnetic ($T_m^*$) and the martensitic ($T_M^*$) transitions.  It is
the end point of a first-order transition line $T^*_I (U^*)$ that, emerging from
the martensitic phase boundary, separates two {\it average} cubic ($\epsilon=0$)
ferromagnetic ($m \neq 0$) phases with different modulation amplitude
($\tilde{\eta}=q_0-q$):  $q^+$ ($\tilde{\eta}<0$) and $q^-$ ($\tilde{\eta}>0$).
Provided the coupling strength is large enough, the IT exists for a limited
range of $K^*$ (below the critical value) which, in turn, depends on $U^*$.

Before ending this section, we would like to show that the magnetoelastic
coupling behind the model under discussion is consistent with the one considered
in the Landau expansion (4), which in turn has been inspired by the experiments.
Let us consider the simplest case of $p=2$ (and $K^*=0$).  Furthermore,
we shall assume that $m_0$ may be approximated by $m_0 \simeq (1-q)m$.  Although
this decoupling (see eq.(17)) is exact only for $T^* \geq T^*_I$, it provides
the first coupling term between $m$ and $q$.  Indeed, from eq. (21) we see that
the magnetoelastic contribution to the internal energy becomes of the form:
\begin{equation}
-2U^* m_0  (m-m_0)  \simeq  -2U^* q(1-q)m^2  
 = \frac{-U^*}{2} m^2 + 2 U^* m^2 \tilde{\eta}^2,
\end{equation}
where $\tilde{\eta}=q_0 - q = \frac{1}{2} - q$.  As it was already discussed at
the end of section (III), the first contribution represents a simple correction
to the Curie temperature, and the second is the magnetoelastic coupling.

\section{Monte Carlo simulations}

In this section we solve numerically the model (\ref{model}) by using Monte
Carlo simulation techniques \cite{MC}.  Our objective is to find system
configurations ($\left \{ \sigma_i \right \},\left \{S_i \right \}$) distributed
according to the canonical ensemble probability.  The corresponding equilibrium
simulations have been carried out using the standard Metropolis algorithm.  The
changes in the $\sigma_i$ and $S_i$ variables are proposed independently and
accepted or rejected according to the single-site transition probability
$W=min\left\{1,e^{-\Delta H^* /T^*} \right \}$.  We have used a 2d-square
lattice with $N$ ($=L^2$) sites subjected to periodic boundary conditions.
Different lattice sizes ranging from $L=20$ to $L=100$ have been studied.  The
unit of time is the Monte Carlo step (MCS) and consists in $N$ attempts of
changing the $\sigma_i$ and $S_i$ variables.  The simulations have been carried
up to $\sim 30 \cdot 10^3$ MCS per site.  Runs have been performed starting from
two initial conditions:  (i) a perfect FMT phase ($\sigma_i=1$,$S_i=1$, $i=1,2,
\cdots N$) and (ii) a perfect FC phase ($\sigma_i=0$,$S_i=1$, $i=1,2, \cdots
N$).  This is very convenient in order to detect metastability and hysteresis
when crossing first-order transition lines.  Notice that, from the mean-field
solution presented in the previous section, we already have an idea of the range
of the space of parameters we have to explore.  Accordingly, we shall fix
$J^*_m=4.0$ and use different values of $U^*<0$.  Concerning the degeneracy
factor we restrict the Monte Carlo simulations to $p=2$.  Nevertheless, we have
verified that other values of $p>2$ render qualitatively similar results.  Most
of the simulations have been performed at fixed values of the model parameters
($U^*$ and $K^*$) and sweeping the temperature $T^*$, but few have been
performed at fixed $T^*$ and sweeping the parameter $K^*$.

The different quantities measured after each MCS are:  the internal energy
${\cal H}^*$, and the order parameters $m$, $\epsilon$ and $q$.  These
quantities have been averaged over $\sim 200$ configurations taken every 100 MCS
and discarding the first $10^4$ MCS for equilibration.  Such averages will be
denoted by $\langle \cdots \rangle$.  We have computed $\langle {\cal H}^*
\rangle$, $\langle |m| \rangle$,$\langle q \rangle$,$\langle |\epsilon|
\rangle$.  Moreover, the specific heat and the susceptibilities associated with
the fluctuations of the order parameters have also been measured:
\begin{eqnarray}
\label{c}
c^* & =  &\left ( \frac{1}{N} \right  ) \frac{\langle {\cal H}^{*2} \rangle -
\langle {\cal H}^* \rangle ^2 }{T^{*2}}\\  \chi_{m}  & = & \frac{\langle m^2
\rangle - \langle |m| \rangle ^2 }{T^*} \\ \chi_{q}  & = & \frac{\langle
q^2 \rangle  - \langle q \rangle ^2   }{T^*} \\ \chi_{\epsilon}  & = &
\frac{\langle \epsilon ^2 \rangle - \langle |\epsilon| \rangle ^2 }{T^*}
\end{eqnarray}
All these definitions correspond to intensive quantities.  In many cases the
specific heat $c^*$ has also been obtained from the numerical derivative $(1/N)
d\langle {\cal H}^* \rangle / dT$.  The agreement between this and the
estimation obtained from eq.  (27) gives confidence that the equilibration times
used are appropriate.  From the behavior of the order parameters, the specific
heat and the corresponding susceptibilities the phase diagram can be obtained.
The phase transitions associated with $\epsilon$, $q$ and $m$ have been determined 
from the location of 
the peaks in either the specific heat or in the corresponding
susceptibilities \cite{fit}.  This method is more accurate than to look for
singularities directly on the behavior of the order parameters.  Moreover, we
have checked whether of not the peaks correspond to a true phase transition by
studying their dependence with increasing the system size $L$.  As an example,
in Fig.  \ref{FIG9} we show the temperature dependence of the specific heat for
$U^*=-3.5$, $K^*=0.15$ and four different system sizes ($L=10,25,50,100$).  The
smooth peak at $T^*\sim 4.3$ does not correspond to a true phase transition
since it does not exhibit scaling behavior.  The second order phase transitions
(at $T^*_M \sim2.25$ and $T^*_m \sim 5.2$) exhibit peaks which shift and become
narrower and higher as one increases the system size $L$ \cite{FSS}.  Besides,
first-order phase transitions ($T^*_I \sim 3.9$ ) exhibit sharp discontinuities
which, although they can also increase, neither shift nor become smoother with
increasing $L$.  In this last case if $L$ is not very large or averages are not
taken for long enough MCS, hysteresis may appear.

Figure \ref{FIG10} shows three sections of the phase diagram as a function of
$K^*$ corresponding to three different values of the coupling parameter:
$U^*=0.0$(a), $U^*=-2.5$(b) and $U^*=-3.5$(c).  We notice that, for each value
of $U^*$, the magnetic ($T^*_m$) and martensitic ($T^*_M$) transitions are
almost independent of $K^*$.  The overall conclusion emerging from Fig.
\ref{FIG10} is that the premartensitic effects are more important the larger the
strength of the magnetoelastic coupling $U^*$ is.  As it was anticipated by the
mean-field calculations, its main effect is the showing up of a critical point
(black diamond), and a first-order transition line (dashed line) separating two
FC phases, $q^+$ and $q^-$, with different values of the modulation amplitude
$\tilde{\eta}$.  Contrarily to the mean-field solution, for $p=2$ , the
$T^*_I(K^*)$ is now bent due to $\langle \sigma_i \sigma_j \rangle$
fluctuations, as will be discussed in the next section.

The second interesting point manifested by the Monte Carlo results is the
existence of large fluctuations close to the critical point which, we stress, do
not correspond to true phase transitions.  These are revealed by anomalies in
the response functions defined in equations (27-30).  In the case of the
specific heat, such anomalies appear in the form of smooth peaks that become
difficult to resolve as we move away from the critical point.  In figure
\ref{FIG11}, which is an enlarged view of figure \ref{FIG10}(c), we denote the
position of such smooth peaks by two dotted lines (with white squares) that,
from the critical point, extend towards both sides of the FC phase.  We have
also indicated the metastability limits associated with the IT (points inside
triangles).  Actually, the position of the IT (denoted by black points along the
dashed line representing $T^*_I(K^*)$) is determined as the middle point of
these limiting lines.  The metastability limit points have been obtained by
performing some runs at constant $T^*$ ($<T^*_{Ic}$) and sweeping $K^*$ (either
increasing $K^*$ from the $q^+$ phase or decreasing $K^*$ starting from the
$q^-$ phase) and some others at constant $K^*$ and increasing the temperature
from the MT phase.  The first-order IT, as will be discussed below, is not found
by decreasing temperature.

Keeping Fig.  \ref{FIG11} in mind, now we shall study the temperature behavior
of the specific heat at constant $K^*$ and $U^*=-3.5$, above and below the
critical point.  The corresponding results are shown in Fig.  \ref{FIG12}.
Different phenomenology may be observed when, increasing the temperature, we
move from the bottom to the top of the figure.

\begin{enumerate}

\item[a)] For $K^*=-0.1$, a martensitic transition and a  magnetic
transition with no sign of an intermediate transition. 

\item[b)] For $K^*=0.12$, a martensitic transition, an anomaly (indicated by 
$\uparrow$) due to the 
proximity of the critical point  and a magnetic 
transition. As we mentioned before, this anomaly is due to fluctuations and  
appears when crossing the dotted line (with 
squares) in Fig. \ref{FIG11}.

\item[c)] For $K^*=0.14, 0.15$ and $0.18$, an additional peak (to the left, also
indicated by $\uparrow$) due to the intermediate transition shows up.  As $K^*$
increases one observes that the fluctuation peak gets smoother (since we move
away from the critical point) while the IT shifts towards lower temperatures.
The entire temperature behavior for $K^*=0.15$ has been previously discussed in
Fig.  \ref{FIG9}.

\item[d)] Finally, for values of $K^*$ even larger ($K^*=0.25$), the anomaly due
to critical fluctuations has almost disappeared and only the peaks associated
with the three phases transitions are clearly revealed:  the martensitic, the
intermediate (or premartensitic) and the magnetic transitions.

\end{enumerate}

It is difficult to be exactly at the critical point (that we have estimated
$K^*_c \simeq 0.130 \pm 0.005$) but it would correspond to the value of $K^*$ at
which the anomaly peak begins to split up.

It is interesting to note that such double peak behavior found around the IT has been 
observed
experimentally in Ni$_2$MnGa.  Figure \ref{FIG13}a shows an example
corresponding to MC simulations with $p=2$, $U^*= -3.5$ and $K^*=0.14$ The two
peaks are found for heating runs only. Due to the metastability of the
first-order intermediate phase transition the transition to the
$q^+$ phase it is not found when cooling.  A similar behaviour is found when 
performing  calorimetric
measurements as it is illustrated in figure \ref{FIG13}b.  The corresponding experimental 
details 
can be found in Ref.\cite{Planes97}.  The lines correspond to thermograms
obtained by heating and cooling (as indicated by the leanning arrows).  Actually, Fig. 
\ref{FIG13}b corresponds to an
enlargement of figure 1 in Ref.  \cite{Planes97} that has been reproduced with
permission of the authors.  Here, additional calorimetric runs are shown in
order to reveal the systematic character of such double peak obtained when
heating.  Indeed,in the original published figure the two peaks are almost 
unobservable and were not considered by the authors who treat both peaks as a
single one.  When comparing Figures \ref{FIG13}a and \ref{FIG13}b one observes
that the highest peak occurs in different order.  We do not have an explanation
for this yet, but it could be related to the calorimeter inertia.

Besides the behavior of the specific heat, it is also instructive to look at the
behavior of the susceptibilities.  Figure \ref{FIG14} shows the evolution of
$\chi_{m}$, $\chi_{q}$ and $\chi_{\epsilon}$ with temperature for $K^*=0.15$ and
$U^*=-3.5$ (this is one of the cases shown in Fig.  \ref{FIG12} and discussed in
point c)).  The smooth peak in the specific heat (marked by $\uparrow$) is
associated only with fluctuations of $q$, while at the IT (large peak) the
discontinuity in $\chi_{q}$ (modulating strain) is accompanied by an increase of
the fluctuations $\chi_{\epsilon}$ (tetragonal homogeneous distortion).

Very recently, measurements of magnetic and transport properties of $NiMnGa$
alloys have been reported \cite{Zuo99}.  Magnetization measurements as a
function of temperature, for small values of the applied field, reveal the
existence of premartensitic anomalies at two different temperatures (separated
$\sim 20^o$).  Only the low temperature anomaly is found when performing
resistivity measurements.  Nevertheless, it has been suggested \cite{Zuo99} that
both correspond to different true pre-martensitic transitions.  In the light of the 
present
results we suggest that the high temperature anomaly could be a signature
of the critical fluctuations.  We agree with the authors that more careful
experimental studies are needed in order to clarify the results.

\section{Discussion and conclusions}

The phase diagram of the model exhibits, qualitatively, the same features in
both mean-field theory and Monte Carlo simulations.  A given value of the
ferromagnetic interaction parameter $J^*_m>0$ determines the distance between
the magnetic $T^*_m$ and the martensitic $T^*_M$ ($<T^*_m)$ transitions.  For
appropriated values of the parameters $K^*$ and $U^*$, the following phases are
found, from high to low temperature:  paramagnetic cubic (PC), ferromagnetic
cubic (FC), ferromagnetic intermediate or premartensitic (FPMT) and
ferromagnetic martensitic (FMT).  The change from the FC to the FPMT phases
occurs below a critical point ($T^*_{Ic}$, $K^*_c$) and it takes place trough a
true phase transition only for $K^* > K^*_c$.  The existence of this
intermediate phase depends on both $K^*$ and $U^*$.  First one needs the
magnetoelastic coupling $U^*$ ($-J^*_m < U^* < 0$) be strong enough.  We obtain
that $T^*_{Ic}$ decreases with $U^*$ whereas $T^*_M$ remains almost unaltered so
that the critical point disappears, on the $T^*_M$ line, well before $U^*$
reaches the value zero (Fig.\ref{FIG6}).  Moreover, provided the $U^*$ is
adequate, the IT exists for a limited range of values of $K^*$($>K^*_c$) across the
first-oder transition line $T^*_I(K^*)$.  The corresponding order parameter is
the modulating strain amplitude and changes from high to low values as one decreases
the temperature across the IT.  According to the theory of the harmonic thermal
vibrations in a crystal, this is consistent with the behavior observed for the
phonon frequency \cite{Zheludev96}.  Moreover, in our results, the IT is
accompanied by a jump in the magnetization, only visible for $p>2$.  Since this
happens in both mean-field and Monte Carlo solutions we conclude that this has
to do with the coupling rather than with the fluctuations.  Experimentally, a
jump in the magnetization has not been detected so far \cite{Webster84,Lvov98}.

Some differences are obtained between the Monte Carlo simulation and the mean
field solutions.  First, in the simulations, the $T^*_I(K^*)$ line always bends 
towards increasing
$K^*$ for $p \geq 2$ whereas the mean-field solution gives, for $p=2$ a perfect
vertical boundary at $K^*=0$.  This is due to $\langle \sigma_i \sigma_j
\rangle$ fluctuations and may be understood as follows.  The internal energy of
a pure $q^-$ phase ($\sigma_i = 0$, $S_i = 1$) is given by (see eq.  (12)):
\begin{equation} 
E^{-} = - K^* 2 N -J_m^* 2 N, 
\end{equation}
while the   energy of  a   pure $q^+$ phase   
($\sigma_i = \pm 1$, at random, $S_i = 1$) is:
\begin{equation} 
E^{+} = - < \sum_{nn} \sigma_i \sigma_j> -J_m^* 2 N.
\end{equation}
In the mean-field approximation the term between $<..>$ is neglected and
therefore the condition $E^{-}= E^{+}$ gives $K^*=0$.  In Monte Carlo
simulations, the fluctuations are present.  We can estimate its value by taking
the results from the standard Ising model and therefore $- < \sum_{nn} \sigma_i
\sigma_j> \simeq E_{Ising}^*(T^*)$.  This function is zero only at $T^*=\infty$.
For $T^*=0$ takes the value $-2 N$ and increases monotonously.  In the Ising
model, at $T^* =T_{c}^* \simeq 2.27$, $E_{Ising}^*=-\sqrt{2}$.  Therefore, the
condition $E^{-}= E^{+}$ gives a transition line at $K^*(T^*) \simeq -
E_{Ising}^*(T^*)/2N$.  It bends towards $K^*>0$ values when $T^*$ decreases.
Remember that in the mean-field approximation a similar behavior is found only
when the degeneracy factor of the 0-state is $p>2$.

Another point illustrated by the Monte Carlo simulations is the important role
played by the fluctuations in describing premartensitic effects.  In figure
\ref{FIG15} we show, schematically, the $K^*-U^*$ section of the phase diagram
as it is obtained from the numerical results presented in the previous section.
The triangle defines the region with a first-order intermediate (premartensitic)
transition and it is limited by the line of critical points (thick line) and the
line where the IT disappears on the MT line (thin
line).  The shadow region denotes the zone of large (critical) fluctuations and
it has been estimated from the anomalies in the specific heat.  It is
interesting that, apart from the region of fluctuations, the mean-field solution
renders a qualitative similar phase diagram (Fig.\ref{FIG8}) although shifted to negative 
values of $K^*$.

Experimentally, the interplay between $U^*$ and $K^*$, required for the IT to
occur, is determined by the composition of the sample.  Indeed, premartensitic
effects in the NiMnGa alloy have only been reported for compositions around the
stoichiometry (Ni$_2$MnGa).  Among them, the most important and common to all
the samples, is a significant softening of the $\frac{1}{3}[110] TA_2$ acoustic
mode with decreasing temperature, accompanying the formation of an intermediate
structure that, while preserving the cubic symmetry, is transversely modulated
\cite{Fritsch94,Cesari97} along the $[110]$ direction with wavevector
$\frac{1}{3}$ .  Moreover, this intermediate structure may appear through a true
phase transition \cite{Planes97,Zheludev96} or not \cite{Stuhr97}.  We point out that 
both
behaviors are compatible with the present results.  In simple words, the
samples showing the IT would fall inside the triangle in Fig.  \ref{FIG15} while
the others not.  More precisely, in the samples showing the IT the strength of
the magnetoelastic coupling ($U^*$) is enough to freeze completely the anomalous
phonon with degree of softness related to $K^*$.  For deeper discussions it is
imperative to have a better understanding about what is behind the composition
dependence giving rise to the different behaviour in the samples.  In
particular, experiments in order to compare the different degree of softness of
the anomalous phonon are required.  Furthermore, it is important to know the
strength of the magnetoelastic coupling.  For this, measurements in both the
ferro and the paramagnetic phases of the temperature behaviour of $\omega_s^2$
are needed.  An study of the characteristics of the kink around the Curie point
would be very helpful.  Finally, we point out that samples in the critical
region (shadow region in Fig \ref{FIG15}) should exhibit a significant
increasing of diffuse scattering when decreasing the temperature below the Curie
point.

The present model renders, independently of the technique used to solve it, a MT
which is continuous for the range of model parameters studied.  This would
be a serious setback in case we where interested in the properties of the MT
itself.  For this, models such as discussed in references \cite{Vasilev99} and
\cite{Lvov98} could be more appropriate.  Here, we have developed a model with
the aim to focus on the study of the IT and related premartensitic effects in
Ni$_2$MnGa.  It is based on the assumption, sustained by the change of the slope
$\frac{d\omega_s^2}{dT}$ at the Curie point, that the magnetism causes the
freezing of the incipiently unstable $\frac{1}{3} [110]$ $TA_2$ phonon and the
splitting from the homogeneous strain.  With this point of view, the
intermediate phase is a precursor of the MT.  On the other hand, in view of the
fact that the actual modulation of the martensitic phase is different from that
of the intermediate phase, some authors have claimed that both phase transitions
have to be regarded as independent \cite{Cesari97PRB}.  If so, additional
coupling terms between the modulation and the homogeneous strains are required
in order to produce a change in the modulation at the martensitic transition.
In this sense, it has been observed that the dip in the $TA_2$ branch shifts
under an external unaxial stress \cite{dip}.  However, more along with the point
of view adopted here, Sthur {\it et al.}  \cite{Stuhr97} have demonstrated that
the $[\frac{1}{3} \frac{1}{3} 0]$ characterizing the intermediate phase becomes
a vector $[0.38, 0.38, 0]$ in the tetragonal phase, very close to the
$[\frac{2}{5} \frac{2}{5} 0]$ expected for a 5-layered modulation.

From the above discussion is follows that, in spite of the appealing results
obtained, the present model needs to be improved in order to reproduce the whole
scenario of the structural transition in Ni$_2$MnGa.  Finally, it is worth
mentioning that, by increasing the strenght of the magnetoelastic coupling
$U^*$, it is possible to extend the present study to samples for which the MT
takes place in a paramagnetic matrix \cite{Chernenko98}

\section{Acknowledgements}

We acknowledge A.  Planes, Ll.Ma\~{n}osa  and E.  Obrad\'o for fruitful  discussions.
This work has received   financial support from CICyT (project  number
MAT98-0315).

\begin{table}
\label{TABLE}
\caption{Identification of the different phases of the model and their 
corresponding
abbreviated notation.  The phase $q^+$ and $q^-$ are both average cubic phases and
undistinguishable above
the  critical  point. For convenience, the $q^+$ phase is also called premartensitic 
FPMT.
Note that $q_0=\frac{2}{p+2}$ takes the value $1/2$ for $p=2$.}
\begin{tabular}{cc|c|c|c} \hline
Name && $\epsilon$& $m$& $q$ \\
\hline
Paramagnetic cubic PC&&$0$&$0$&-\\
\hline
&pure cubic&&&$q_0$\\
\cline{2-2} \cline{5-5}
Ferromagnetic cubic FC &average cubic ($q^-$)&$0$&$\neq 0$&$< q_0$\\
\cline{2-2} \cline{5-5}
&Premartensitic FPMT ($q^+$)&&&$>q_0$\\
\hline
Ferromagnetic martensitic FMT&&$\neq 0$&$\neq 0$&$> q_0$\\ 
\hline
\end{tabular}
\end{table}

\begin{figure} \psboxto(0.8\textwidth;0cm){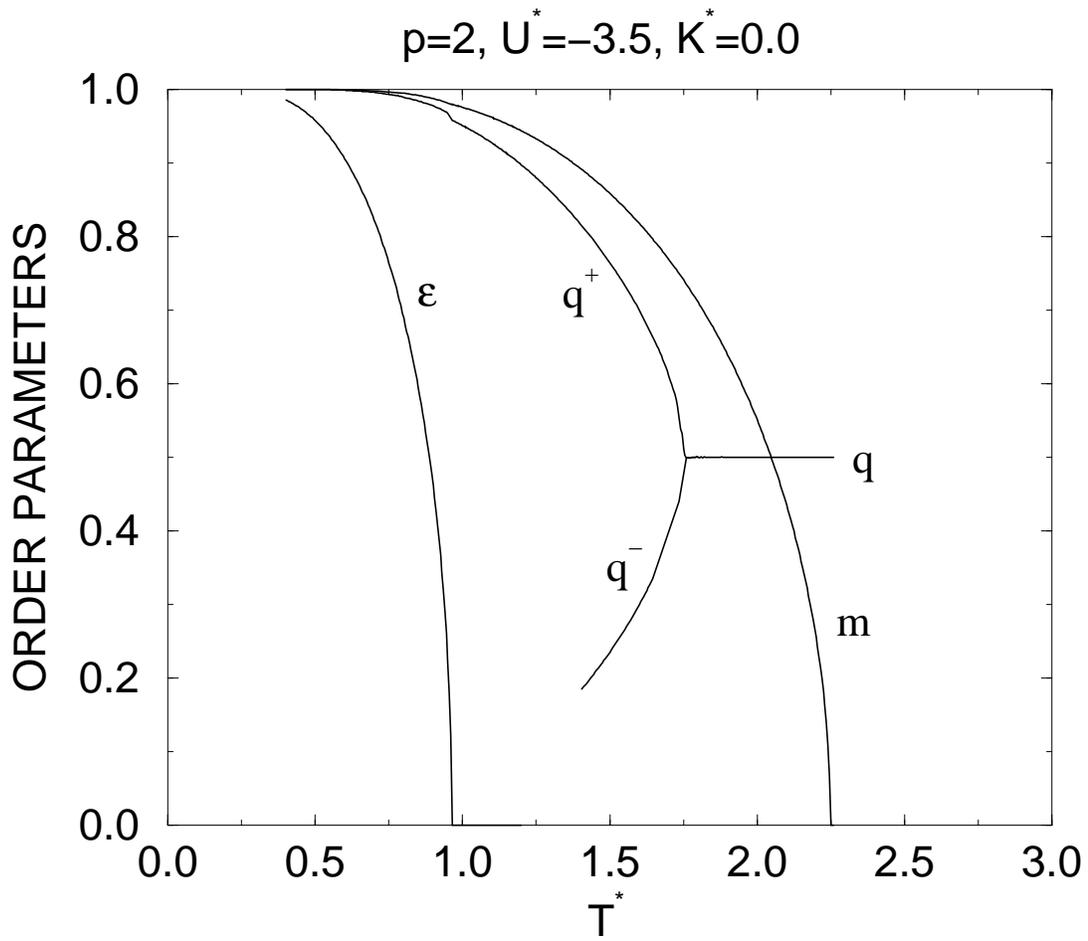} \caption{Mean-field 
temperature evolution of the order parameters for $J^*_m=4.0$, $U^*=-3.50$, 
$p=2$, along the coexistence line, that in this case ($p=2$) corresponds to $K^*=0$.}  
\label{FIG1} \end{figure}

\begin{figure} \psboxto(0.8\textwidth;0cm){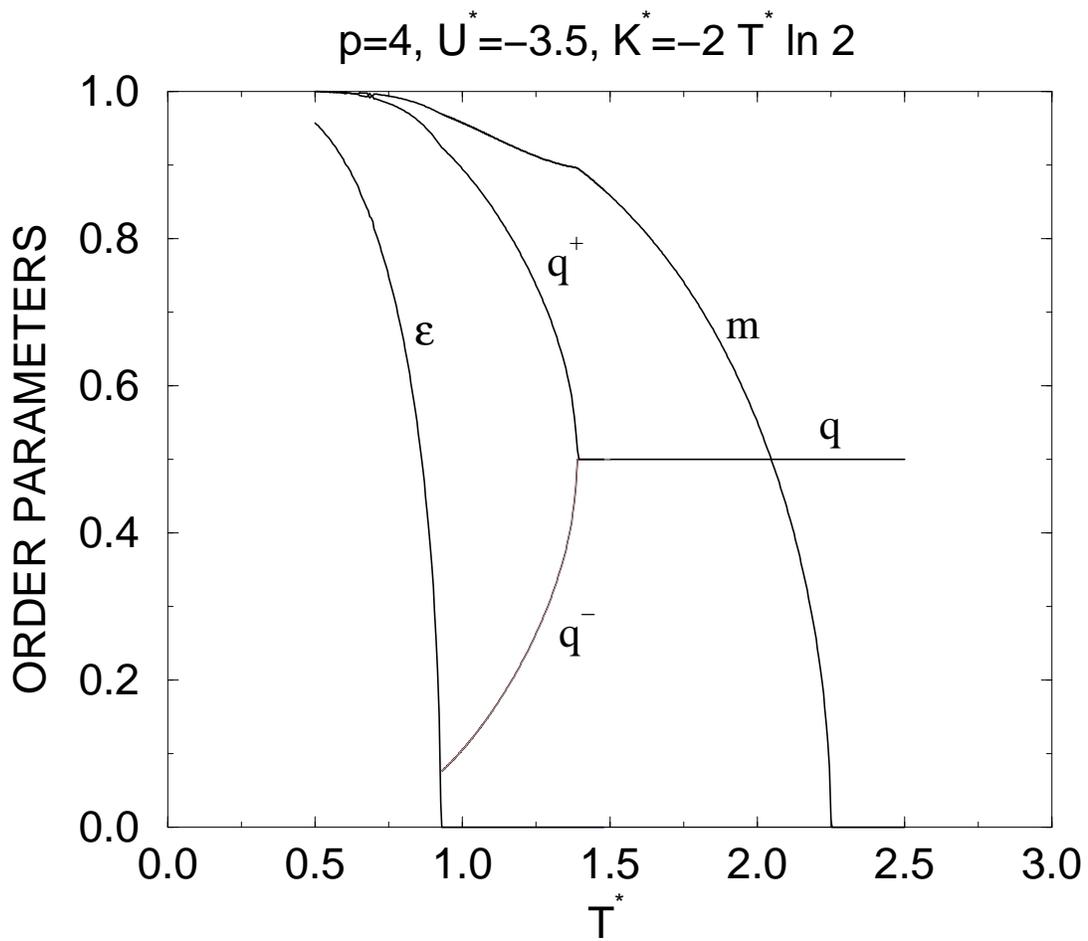} \caption{Mean-field, 
temperature evolution of the order parameters for $J^*_m=4.0$, $U^*=-3.50$, 
$p=4$ along the coexistence line $K^*(T^*)=-2T^* \ln(p/2)$.}  \label{FIG2}
\end{figure}

\begin{figure} \psboxto(0.8\textwidth;0cm){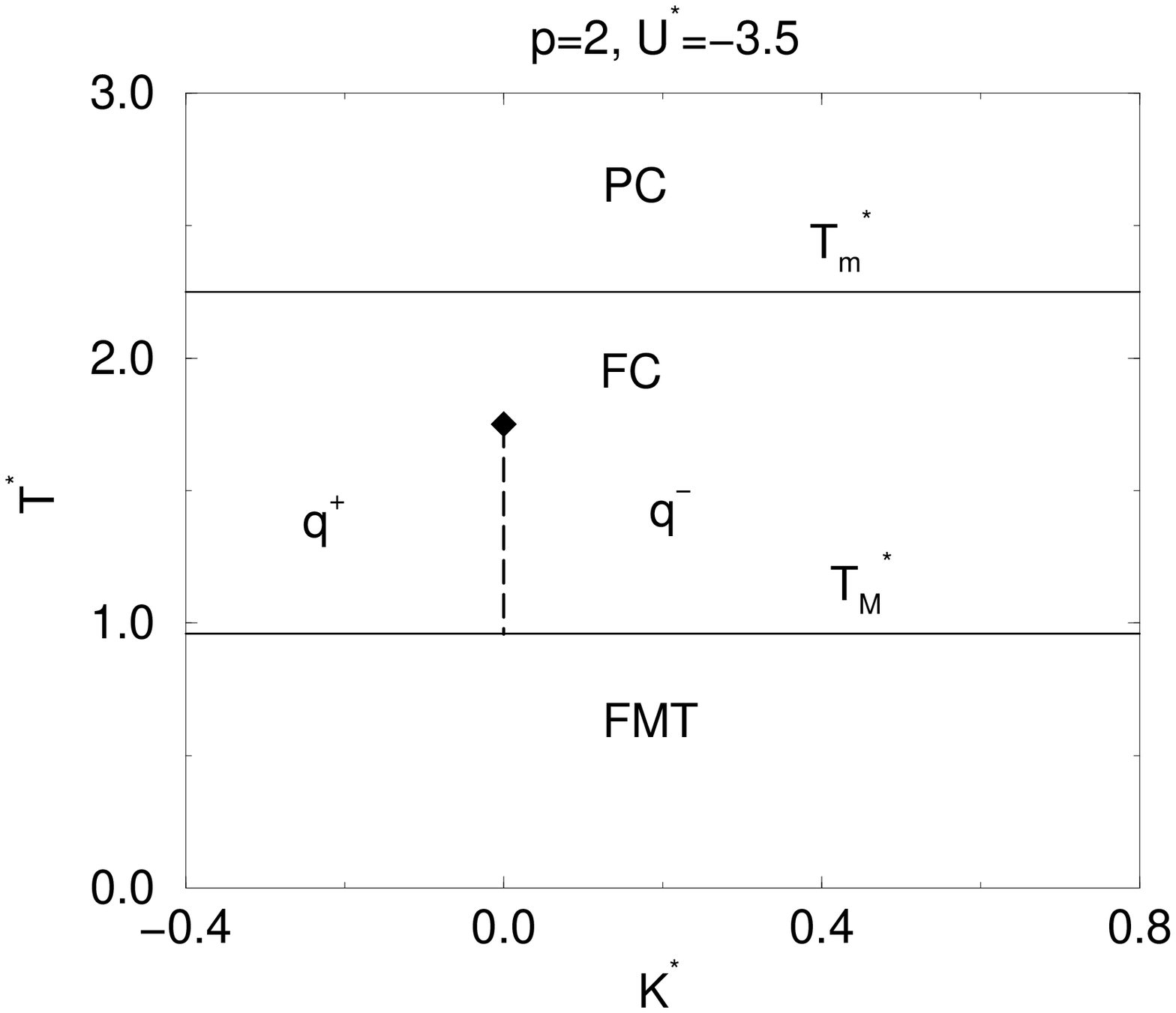} \caption{Section of the
phase diagram obtained by mean-field techniques for $J^*_m=4.0$, $U^*=-3.5$ and 
$p=2$. 
Continuous lines stand for second-order phase transitions and dashed lines for
first-order ones. The black diamond shows the location of the critical point. 
The labels indicate the different phases as explained in Table I.}  
\label{FIG3}
\end{figure}

\begin{figure} \psboxto(0.8\textwidth;0cm) {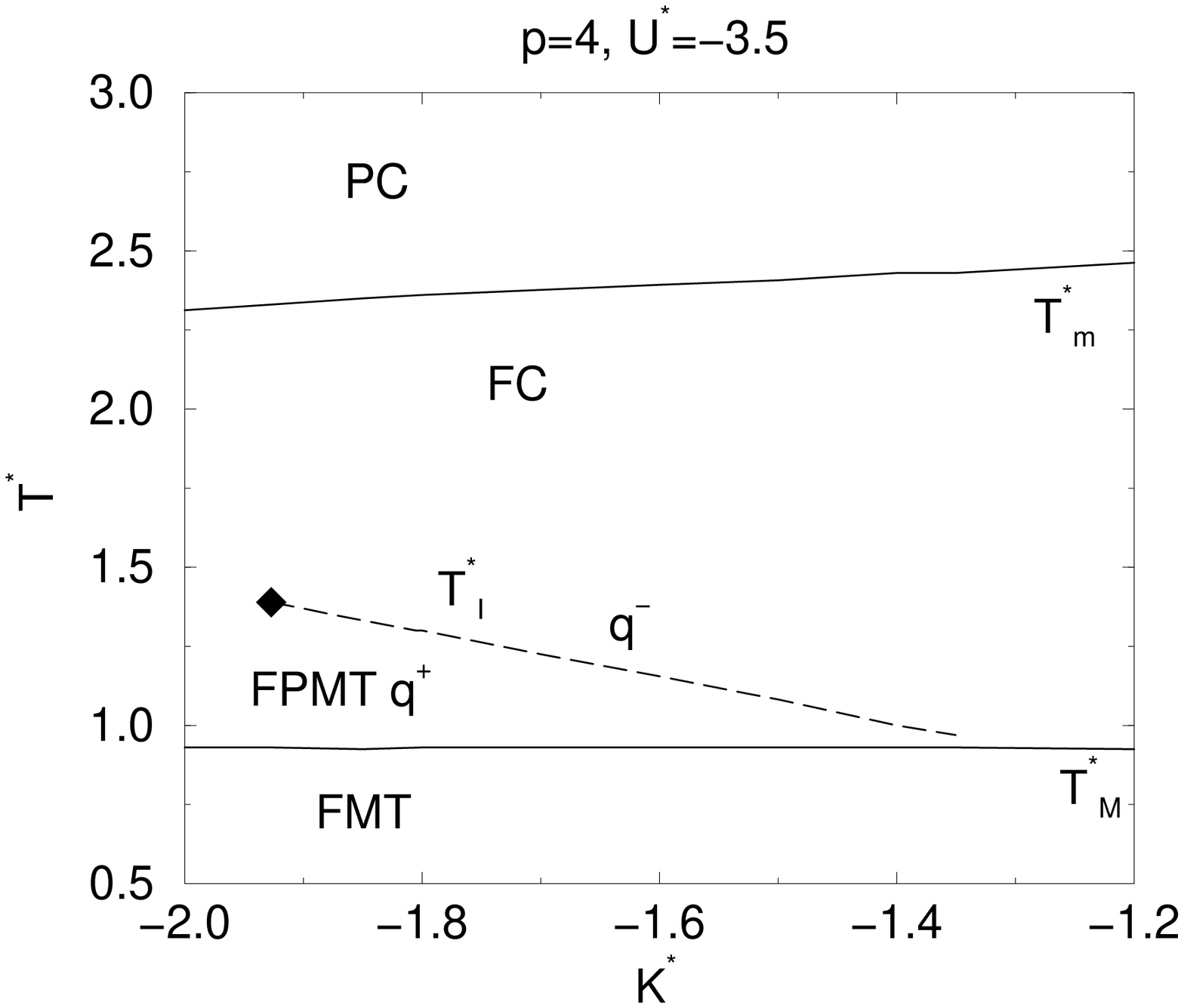} \caption{Section of the
phase diagram obtained by mean-field calculations for $J^*_m=4.0$, $U^*=-3.5$ 
and 
$p=4$. Continuous lines stand for second-order phase transitions while 
dashed lines for
first-order ones. The black diamond shows the location of the critical point. 
The labels indicate the different phases as explained in Table I.}
\label{FIG4}
\end{figure}

\begin{figure} \psboxto(0.8\textwidth;0cm){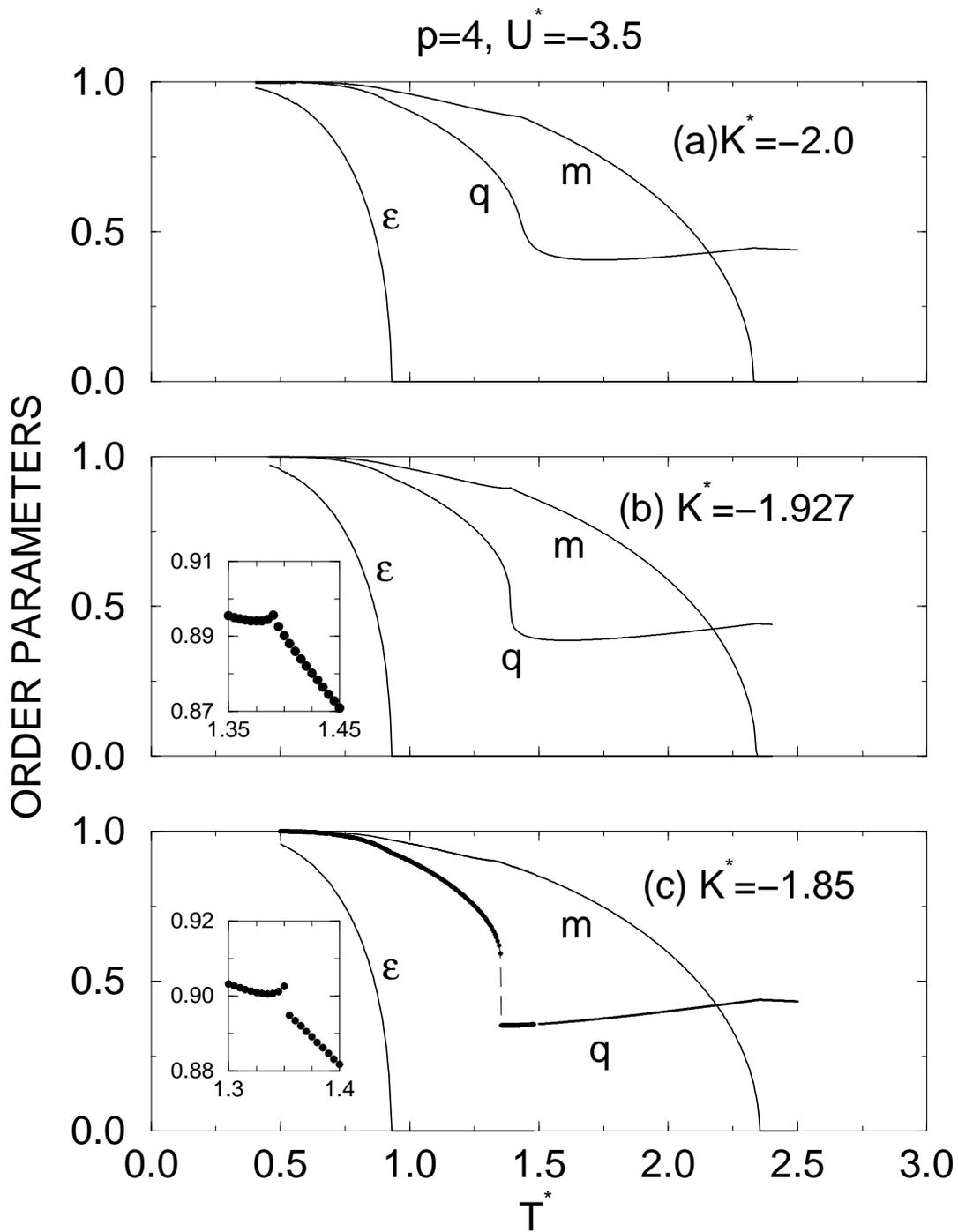} \caption{Mean-field 
temperature evolution of the order parameters for $J^*_m=4.0$, $U^*=-3.50$, 
$p=4$ and three different values of $K^*$: (a) $K^*=-2.0 < K^*_c$, (b) 
$K^*=K^*_c$ and (c) $K^*=-1.85 <K^*_c$. The insets show the detail of the
behavior of the magnetization $m$ at the intermediate transition.}  
\label{FIG5}
\end{figure}

\begin{figure} \psboxto(0.8\textwidth;0cm){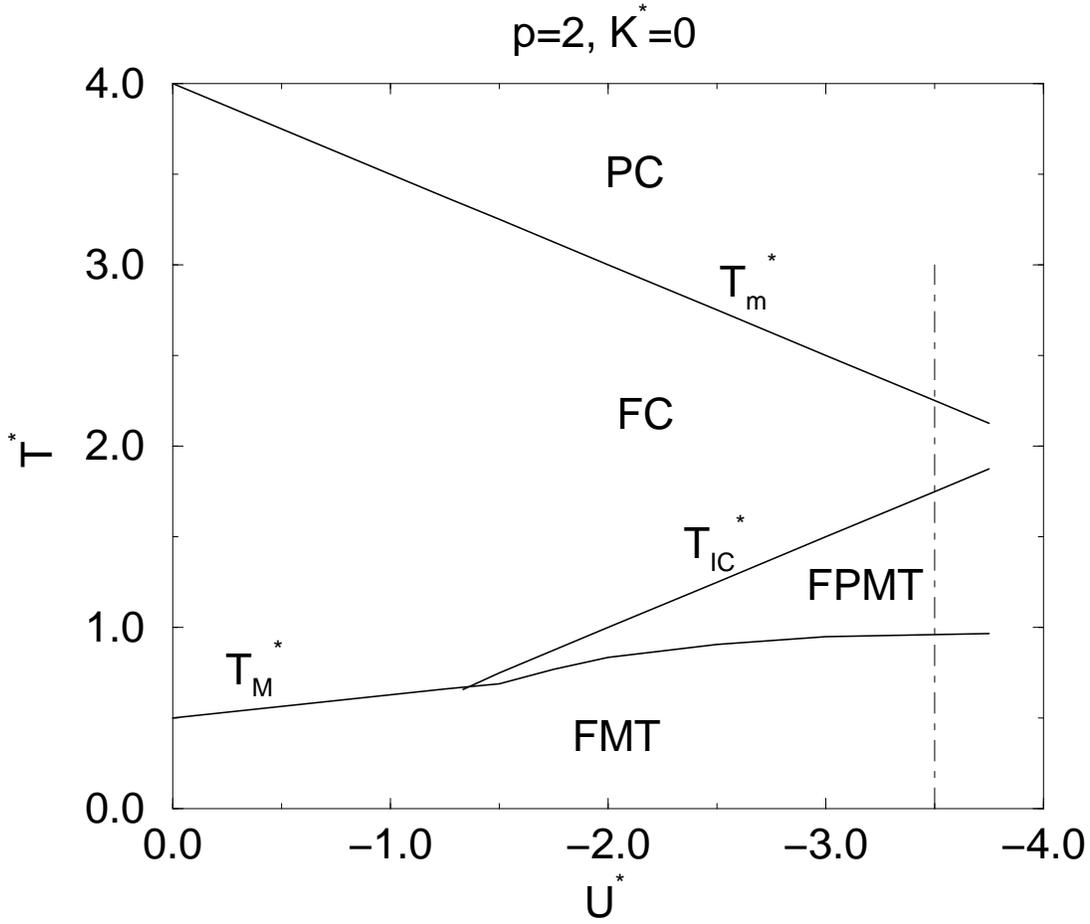} \caption{$U^*$-$T^*$ 
section obtained from mean-field calculations for $J^*_m=4.0$, $p=2$ and 
$K^*=0$. Continuous lines stand for second-order phase transitions. The 
labels indicate the different phases as explained in Table I. The intermediate
region indicated by FPMT is a phase separation region with coexistence of the 
two phases with $q^+$ and $q^-$. The thin dashed line indicates the section 
of the phase diagram shown previously in Fig. 3.}  \label{FIG6}
\end{figure}

\begin{figure} \psboxto(0.8\textwidth;0cm) {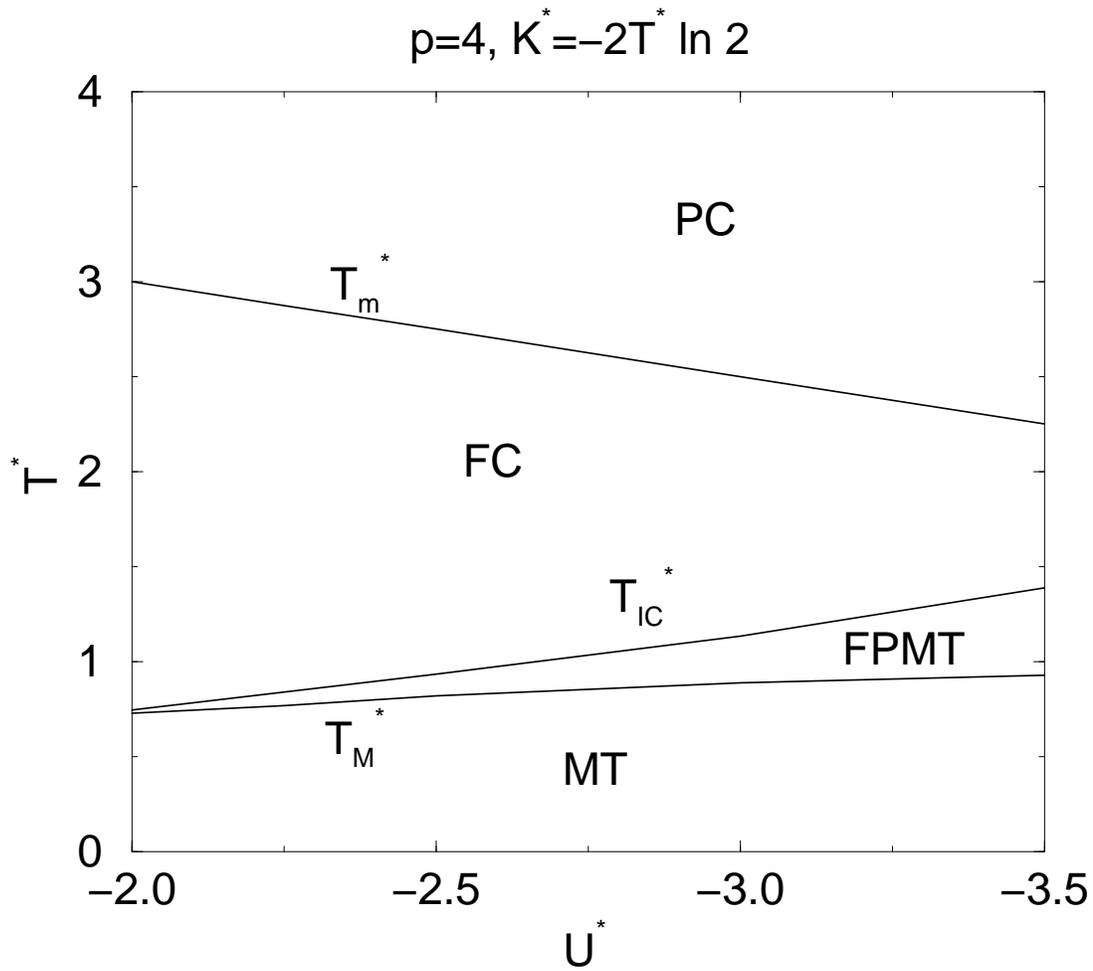} \caption{$U^*$-$T^*$ 
section obtained from mean-field calculations for $J^*_m=4.0$ and $p=4$.  
For each value of $U^*$, $K^*$ is selected in order to find the critical point. 
Continuous lines stand for second-order phase transitions. The 
labels indicate the different phases as explained in Table I. The intermediate
region indicated by FPMT is a phase separation region with coexistence of the 
two phases with $q^+$ and $q^-$.}
\label{FIG7}
\end{figure}

\begin{figure} \psboxto(0.8\textwidth;0cm) {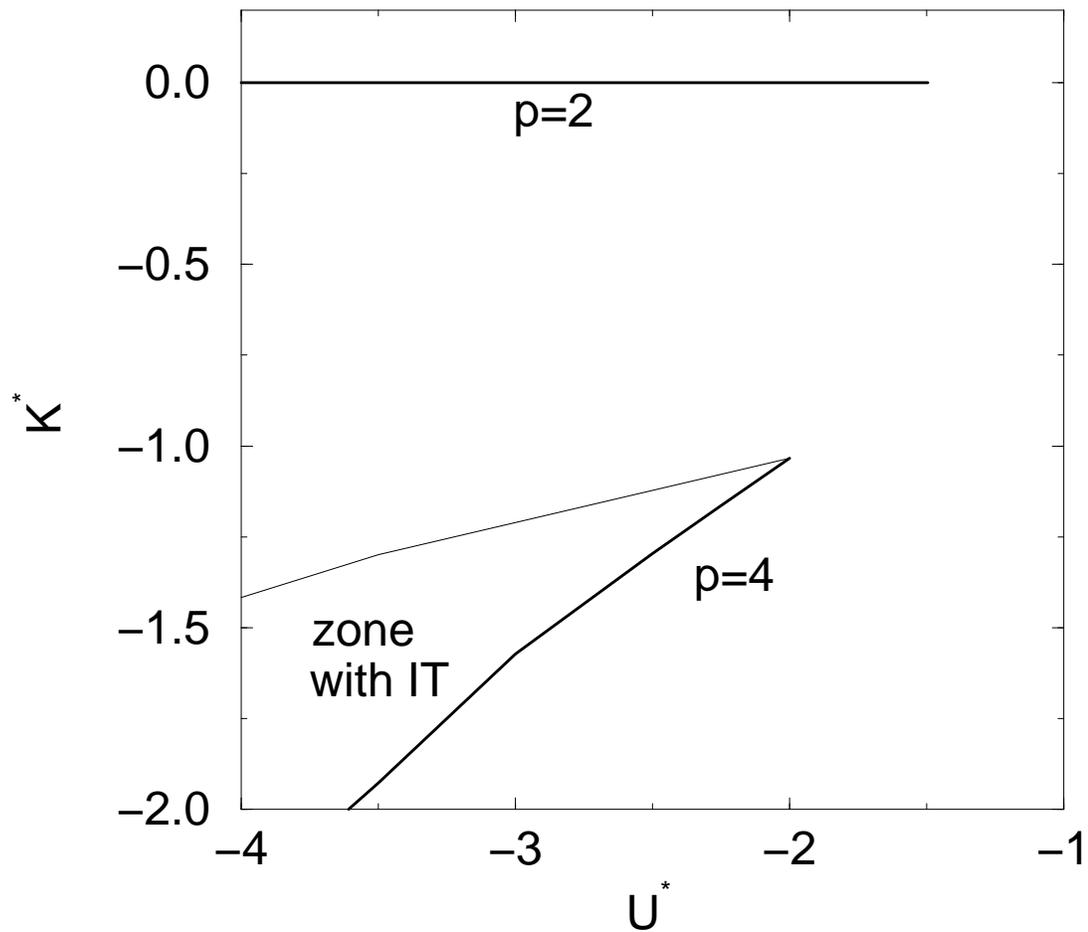} \caption{$U^*$-$K^*$ 
section of the phase diagram obtained from mean-field calculations for
$p=2$ and   $p=4$.   The location of the critical  points  are  shown with
a thick continuous line. The thin line is estimated from the value of $K^*$ at which 
the IT and the MT meet. The region  inside the triangle  denotes the range of 
the model
parameters    for which  there  exists the   first-order  intermediate  
transition (IT) in the case of $p=4$.}
\label{FIG8}
\end{figure}

\begin{figure}
\psboxto(0.8\textwidth;0cm){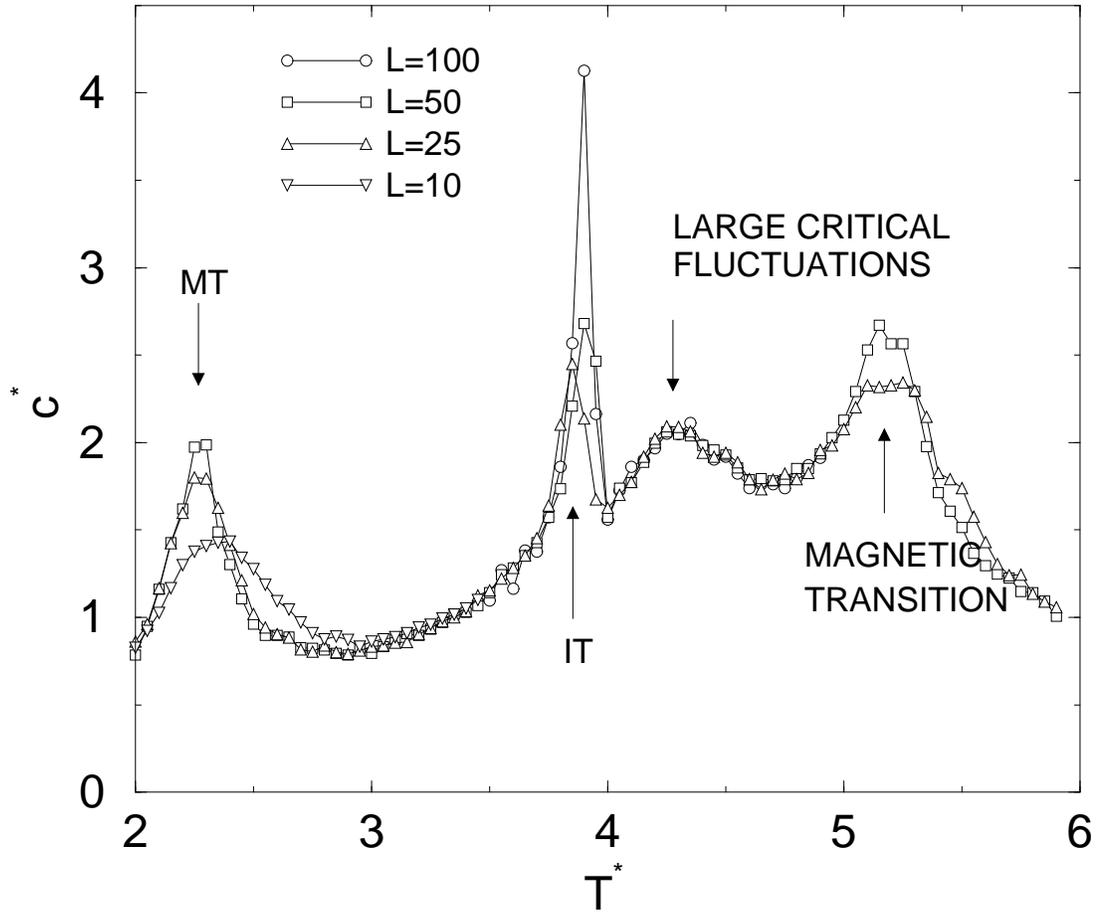} \caption{Example of 
the  scaling with the  system  size $L$ (as indicated by the legend) 
of the  specific heat $c^*$. Note that the MT and the magnetic transition 
exhibit  shifting and increasing peaks, the intermediate transition (IT) 
peaks also increase 
but 
do
not show a clear shift tendency and the peak corresponding to
critical fluctuations does not change at all with $L$.}  
\label{FIG9} \end{figure}

\begin{figure} \psboxto(0.7\textwidth;0cm){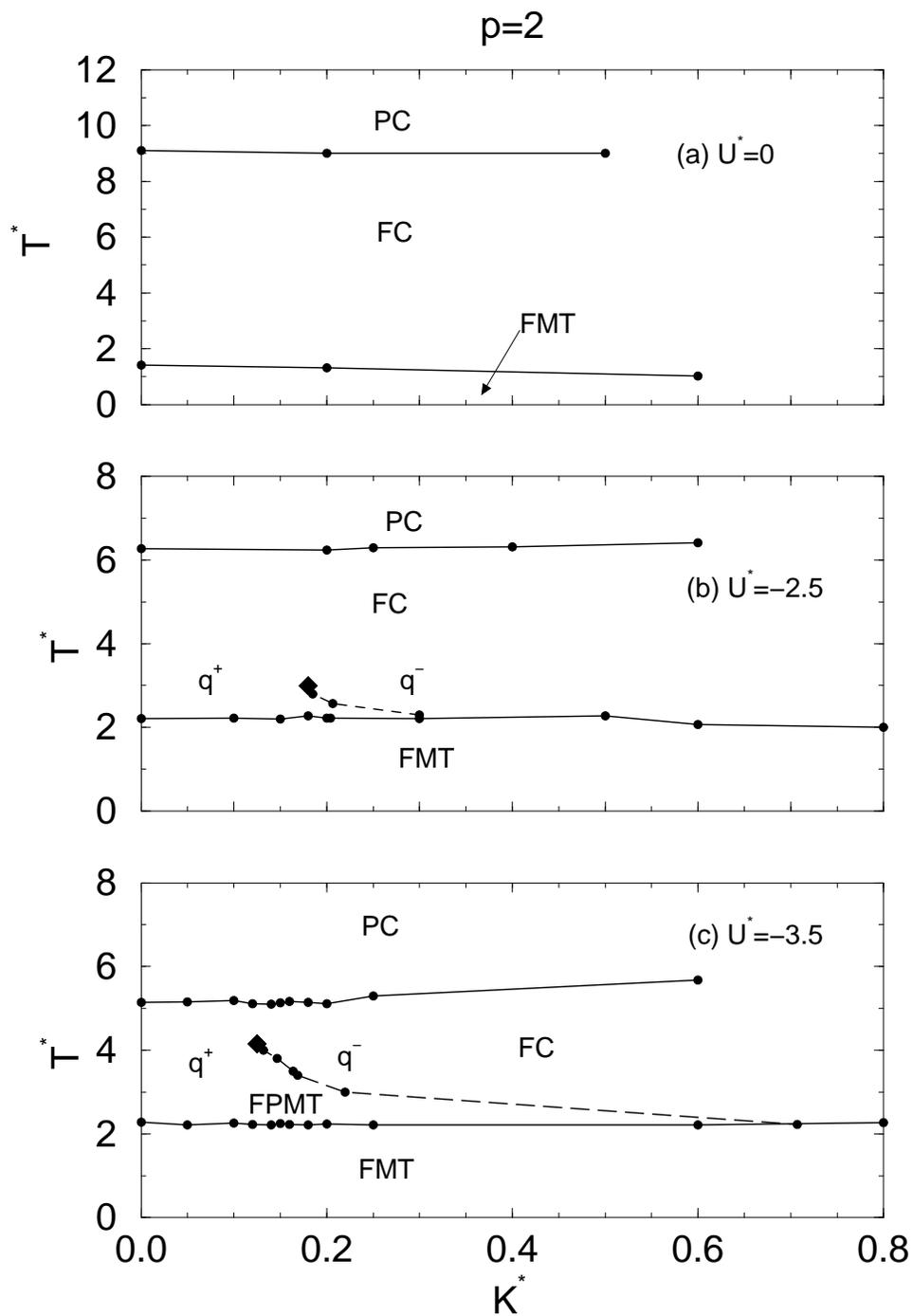} \caption{Sections 
of the phase diagram obtained by MC simulation with $J_m^*=4$  and 
(a) $U^*=0$, (b)$U^*=-2.5$ and (c) $U^*=-3.5$. Dots indicate the actual 
numerical 
data. Lines are guide to the eyes, indicating second-order
transitions (continuous) and first-order transitions (dashed). The 
(approximate) 
position of the critical point is shown by a black diamond. The labels 
indicate 
the different phases as explained in Table I.}  \label{FIG10} \end{figure}

\begin{figure} \psboxto(0.8\textwidth;0cm){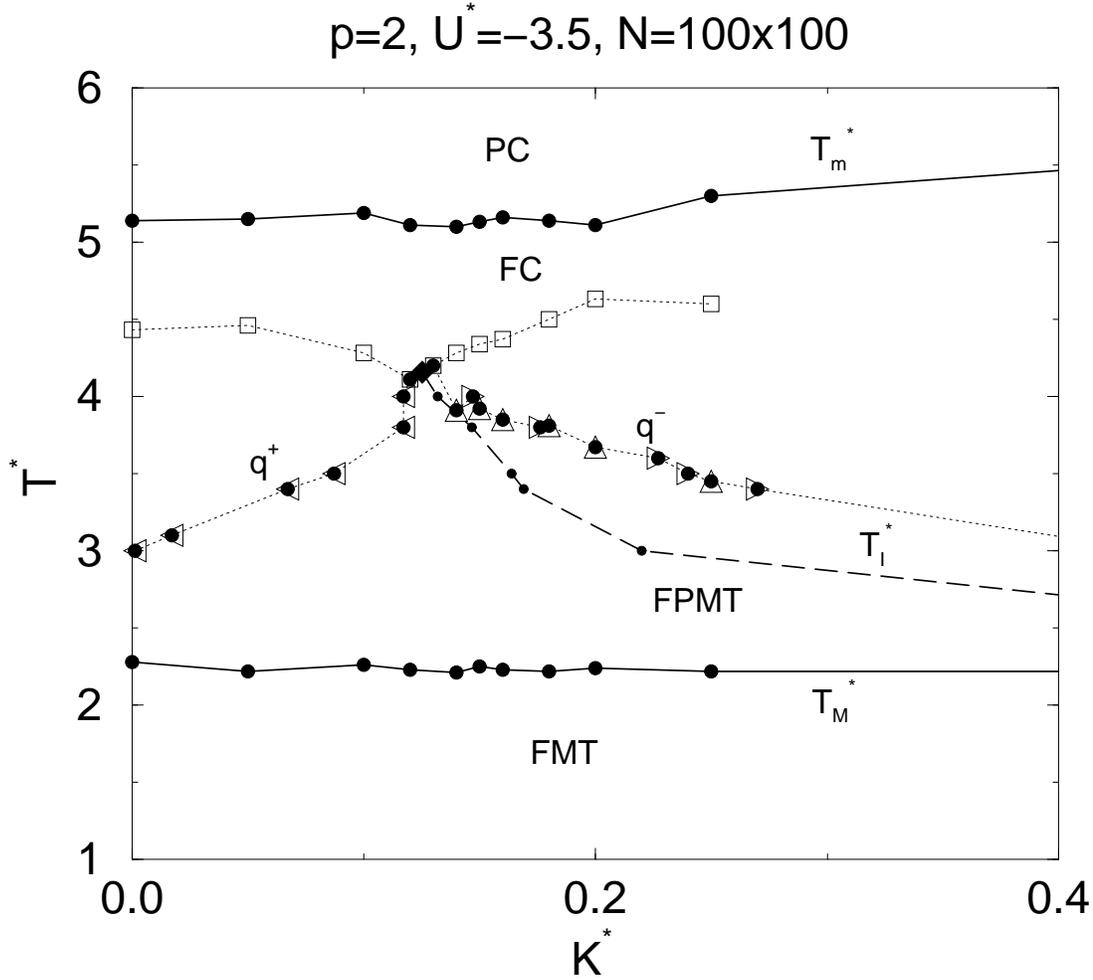} \caption{Detailed 
section of the phase diagram (Fig. 10(c)) obtained  by MC simulation with  
$U^*=-3.5$ and $J_m=4$, showing the
details  of  the metastability  regions  and the   fluctuations in the
neighborhood of the critical point. Symbols and lines have the same meaning as 
in Fig. 10. Besides, we have indicated the position of the large fluctuation 
peaks with squares and the metastability limits of the first-order transition
 lines with points inside triangles. The orientation of the triangles indicates the direction
 of the MC simulation runs performed in order to locate each metastability 
limit.
 Note that we have performed increasing $T^*$ runs ($\triangle$) and 
increasing ($\rhd$) and decreasing ($\lhd$) $K^*$ runs.
Dotted lines are guides to the eyes}  \label{FIG11} \end{figure}

\begin{figure} \psboxto(0.8\textwidth;0cm){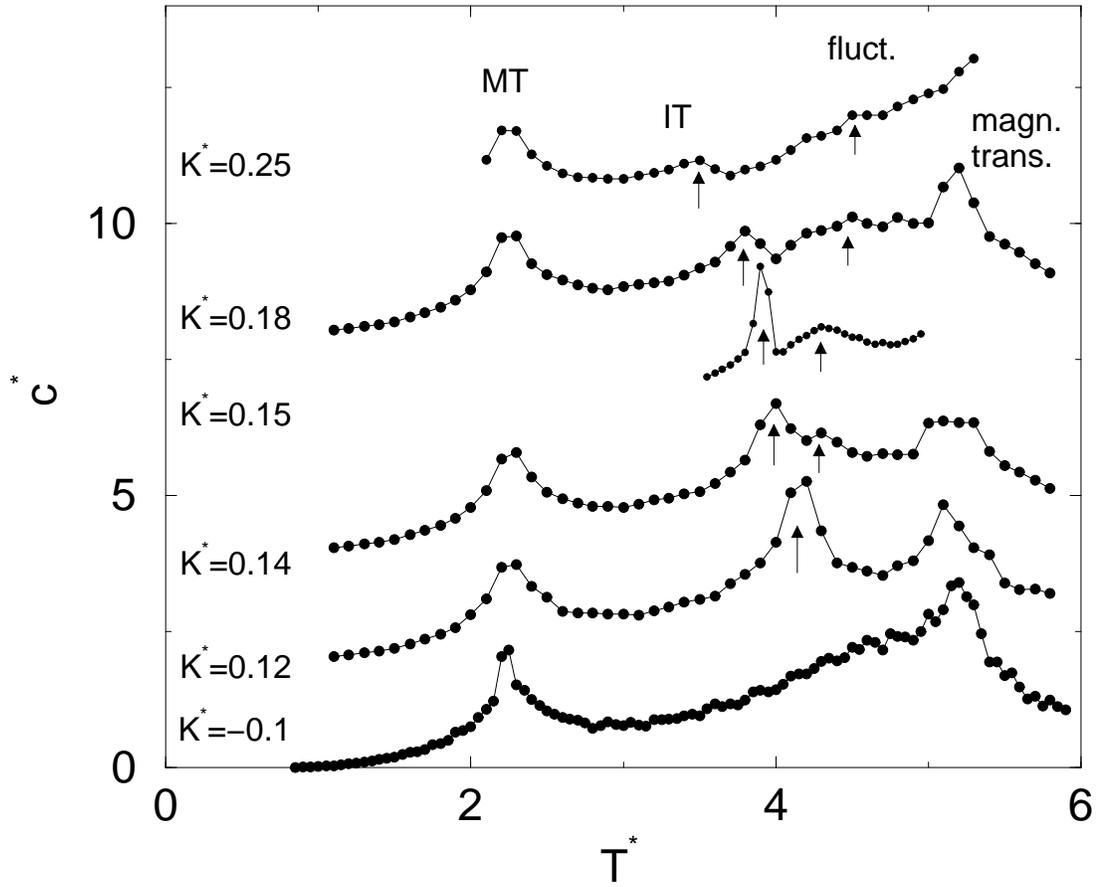} \caption{Specific heat
evolution,  corresponding to $U^*=-3.5$  and different values of $K^*$
as indicated. The arrows on each plot indicate the peaks corresponding to
the IT and the large fluctuations due to the closeness of the critical point.} 
\label{FIG12} \end{figure}

\begin{figure} \psboxto(0.8\textwidth;0cm){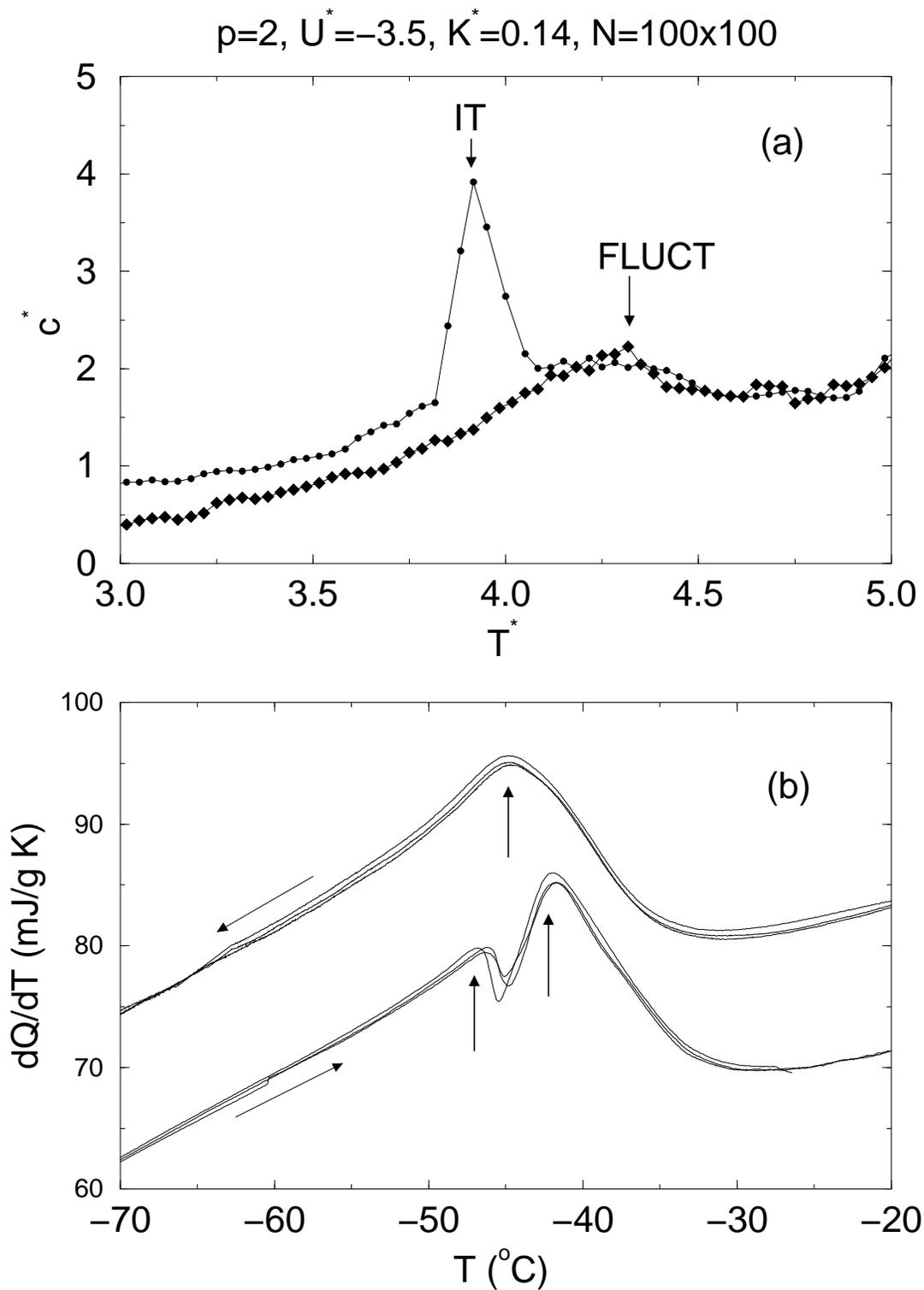} \caption{Details of the
specific heat evolution at the premartensitic phase transition showing
the double peak effect and hysteresis. (a)  Results from MC simulation
of the present model corresponding to $U^*=-3.5$ and$K^*=0.14$ and (b)
experimental data extracted from Ref. 11} \label{FIG13} \end{figure}

\begin{figure} \psboxto(0.8\textwidth;0cm){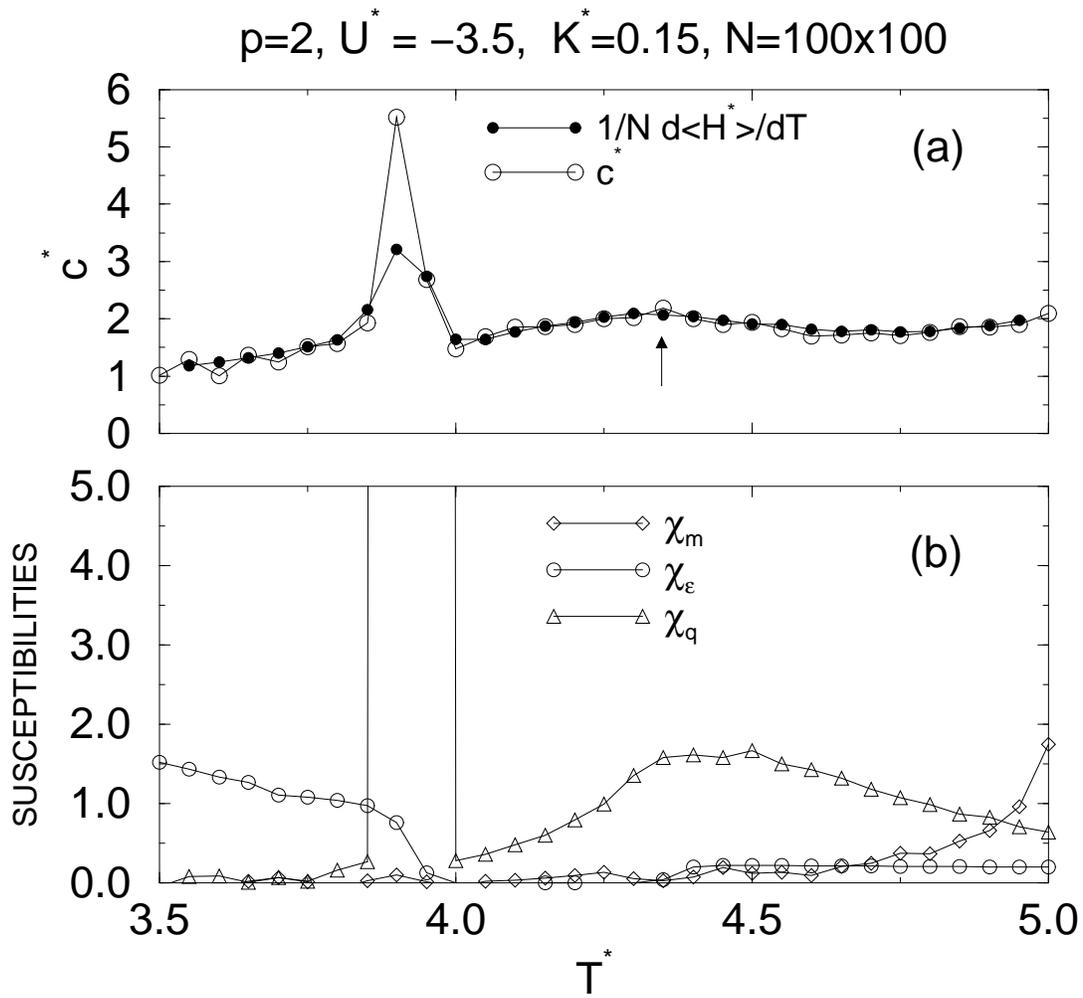} \caption{Comparison of 
the specific heat behavior (a) and the susceptibilities (b). In (a),  the two 
curves correspond to the estimation of $c^*$ from the
derivative of the energy ($\bullet$) and from the fluctuations as defined in the 
text ($\circ$).
In (b) the three different curves correspond to the fluctuations of the
three order parameters $\epsilon$, $q$ and $m$, as indicated by the legend. }  
\label{FIG14} \end{figure}

\begin{figure} \psboxto(0.8\textwidth;0cm){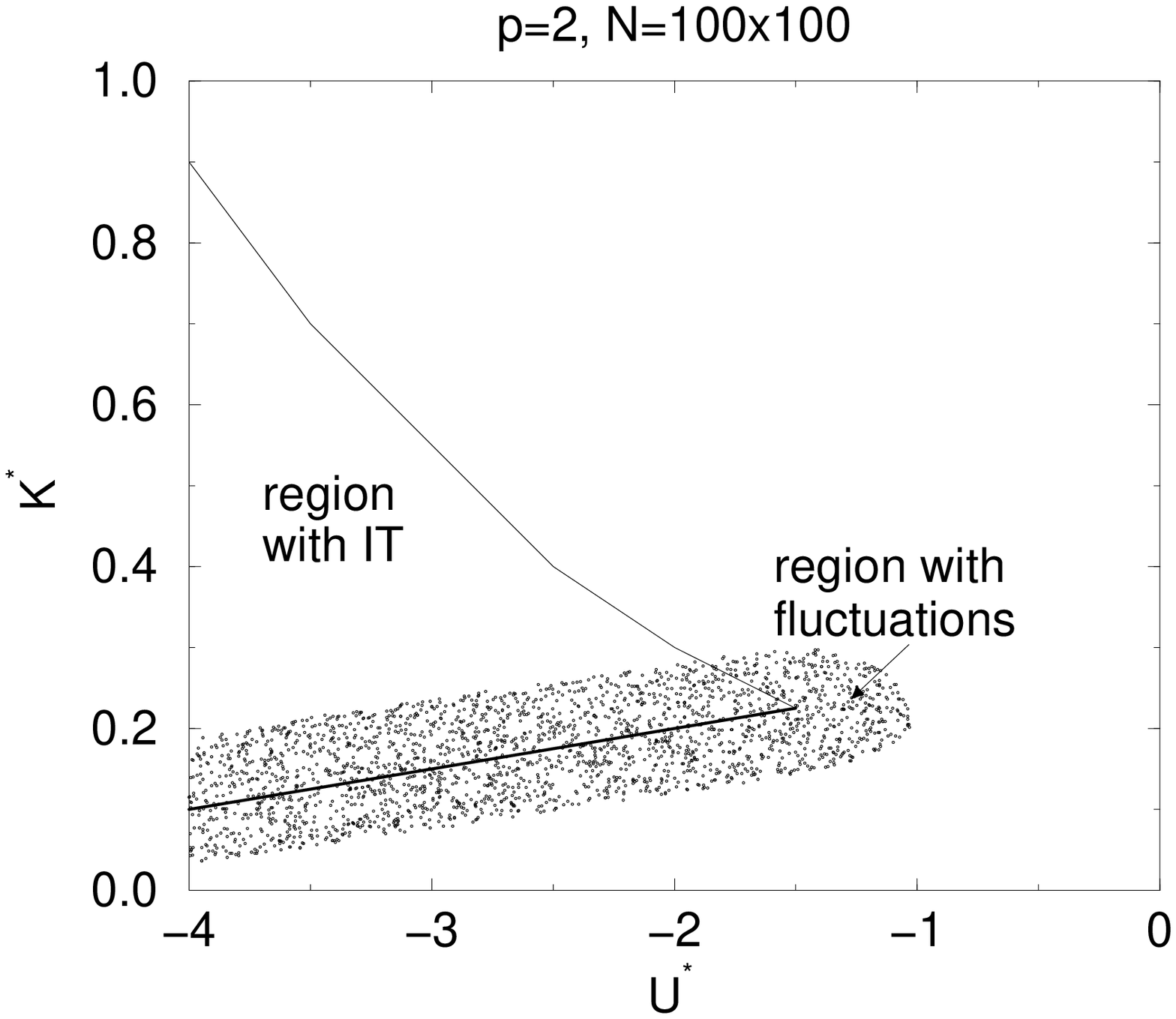} \caption{Space parameters
$U^* - K^*$ obtained with MC simulations, indicating the regions with IT and 
with large fluctuations.  The thick line indicates points with true 
critical 
behavior.}  \label{FIG15} \end{figure}



\begin{references}
\bibitem{PMT} This  intermediate  phase is also  called premartensitic
phase.
\bibitem{comment} In  ref.\protect{\cite{Lindgard86}}  the coefficient
to the internal strain  was in order  to emphasize the direct relation
to the soft mode frequency, for simplicity just written as $\omega_N$,
whereas the more  correct form of  course is  in terms  of the squared
frequency    as  used here:      $m \omega_N^2$.
\bibitem{mdirec} The  easy  direction  of magnetization at    very low
temperatures is parallel  to the $\langle 111  \rangle$ of  the $L2_1$
phase.   Nevertheless,  associated  with  the  martensitic   transition
exhibited al low temperatures, this direction may be different.
\bibitem{magnetostriction}  The  $\sigma_i \sigma_j  S_i S_j$ term
would correspond to the magnetostriction coupling  $\epsilon^2 m^2$, which is already 
effectively included in eq.(2).
\bibitem{fit}   In the first case we have fitted a parabolic peak while in the
second one will need a 4rth order polynomial fit which always involves larger
error bars.
\bibitem{FSS} We have verified that the peaks corresponding to second
order transitions scale according to  the standard finite size scaling
assumptions.  Nevertheless, the detail of  such analysis is out of the
scope of this paper.
\end{references}
\end{document}